\documentclass{jpp}
\usepackage{graphicx}
\usepackage{float}
\usepackage[utf8]{inputenc}
\usepackage[T1]{fontenc}
\usepackage{mathptmx}
\usepackage{amsmath}
\usepackage{etoolbox}
\usepackage{bbold}
\usepackage{bbm}
\usepackage{txfonts}
\usepackage[dvipsnames]{xcolor}
\usepackage[colorlinks=true,linkcolor=red,citecolor=blue,urlcolor=violet]{hyperref}
\graphicspath{ {images/} } 
\usepackage{multirow}
\usepackage{soul}
\usepackage{graphicx}

\shortauthor{D. L. Schröder et al.}

\title{On the modeling of oblique firehose instabilities in regularized 
Kappa plasmas using ALPS}

\author{D. L. Schröder\aff{1}
  \corresp{\email{dustin.schroeder@rub.de}},
  M. Lazar \aff{2,1},
  H. Fichtner\aff{1},
  K.G. Klein\aff{3}
 \and D. Verscharen\aff{4}}

\affiliation{\aff{1}Lehrstuhl für Theoretische Physik IV: Plasma-Astroteilchenphysik, Ruhr-Universität Bochum, D-44780 Bochum, Germany
\aff{2}Centre for mathematical Plasma-Astrophysics, KU Leuven, 3001 Leuven, Belgium
\aff{4}Mullard Space Science Laboratory, University College London, Dorking RH5 6NT, UK
\aff{3}Department of Planetary Science, University of Arizona, Tucson, USA}

\begin{document}

\maketitle

\begin{abstract}
In-situ measurements in space plasmas indicate that the velocity distributions of charged particles are not in thermal equilibrium, deviating from a standard Maxwellian mainly due to anisotropies and suprathermal populations which enhance high-energy tails. 
Although the Standard $\kappa$-Distribution (SKD) is well established in modeling these non-equilibrium distributions, its application is often viewed controversially due to certain unphysical implications, in particular divergent velocity moments.
To address these issues, the Regularized $\kappa$-Distribution (RKD) was introduced. 
Such advanced, in general anisotropic RKDs are invoked here for the first time to investigate oblique firehose instabilities, including those induced by the temperature anisotropy of electrons and protons (the dominant species in space plasmas).
In weakly collisional plasmas, both of these instabilities are expected to play significant roles in the self-regulation of the macroscopic properties of space plasmas (such as the expanding solar wind) reported by observations.
Is not yet possible to resort to a general dispersion tensor related to RKD plasmas (whose derivation is still a challenge due to the complexity of these models), instead the new generation Arbitrary Linear Plasma Solver (ALPS) is exploited here.
The unstable solutions obtained for the already established  Maxwellian and SKDs successfully validate the capability of ALPS.
For RKDs, the instabilities confirm the stimulating effect of suprathermal populations, with lower $\kappa$ values generally enhancing the firehose growth rates. 
For the oblique electron firehose instability, RKDs substantially modify the competition between periodic and aperiodic branches only at intermediate angles and can sustain significantly increased growth rates of especially aperiodic modes at highly oblique propagation beyond both the Maxwellian unstable regime and the one that is accessible with SKDs.

\end{abstract}

\section{Motivation and Theory} \label{sec:level1}

The dynamics of weakly or non-collisional plasmas, such as the rarefied and hot systems throughout the heliosphere, astrophysical objects, and fusion devices, is largely     determined by plasma waves. 
A relevant example are the wave instabilities (locally) self-generated by plasma particles with kinetic anisotropies in velocity space \citep[e.g.,][]{Gary-1993, Shaaban-etal-2021}.
Kinetic instabilities convert the free energy of anisotropic distributions with various deviations from thermal equilibrium, e.g., particle beams, temperature anisotropy, and suprathermal or non-Maxwellian populations. 
Quasi-stable states of plasma particles in the solar wind and the Earth's magnetosphere, mostly electrons and protons, show a clear tendency of confinement below the anisotropy thresholds of these instabilities in parameter space \citep{Kasper-etal-2002, Samsonov-etal-2007, Stverak-etal-2008, Bercic-etal-2019, Klein-etal-2019ApJ}. 
In-situ observations reveal enhanced wave fluctuations associated with electromagnetic (EM) cyclotron and mirror-mode instabilities triggered by a temperature anisotropy $T_\perp > T_\parallel$, and to firehose instabilities driven by an opposite anisotropy $T_\perp < T_\parallel$  \citep[where $\parallel,\perp$ are directions with respect to the magnetic field,][]{Bale-etal-2009, Wilson2013JGRA, Zhao-etal-2019}. 
A definitive answer regarding the role played by these instabilities in constraining the macroscopic properties of plasma populations with complementary fluid modeling, is however conditioned by an adequate modeling and interpretation of their distributions in the velocity space.

\subsection{Velocity distribution functions}

Quasi-thermal particle populations in magnetized plasmas are most often described by anisotropic Maxwellian distributions. 
The non-drifting bi-Maxwellian distribution function $f_M$ is given by
\begin{equation}\label{fmaxwell}
f_M(v_{\|}, v_{\perp})= N_M \exp{\left(- \frac{v_{\|}^2}{\theta_{M \|}^2}- \frac{v_{\perp}^2}{\theta_{M,\perp}^2}\right)},
\end{equation}
where $v_{\|}$ and $v_{\perp}$ represent velocity components parallel and perpendicular to the background magnetic field. 
The thermal speeds are $\theta_{M \|, \perp} = \sqrt{2k_B T^M_{\|, \perp}/m_j}$ with the Maxwellian temperature $ T^M_{\|, \perp}$, the Boltzmann constant $k_B$, the particle rest mass $m_j$ of species $j$, and the normalization constant reads $N_M=1/(\pi^{3/2} \theta_{M,\|} \theta_{M, \perp}^2 )$.

In situ observations in space plasmas also reveal suprathermal particle populations which manifest as high-energy tails of the observed distributions, that cannot be captured by Maxwellian models alone. 
These suprathermal tails are often described using $\kappa$-distributions, originally introduced by \citet{Olbert1968} and widely applied in studying kinetic instabilities  \citep{Meneses_POP2018, Shaaban-etal-2021, Lopez-etal-2021}.
The standard bi-$\kappa$-distribution (SKD) can be written as \citep{Lazar_Fichtner_Yoon_2016}
\begin{equation}\label{f_skd}
f_{\mathrm{SKD}}\left(v_{\|}, v_{\perp}, \kappa \right)=N_{\mathrm{SKD}}\left(1+\frac{v_{\|}^2}{\kappa \theta_{\|}^2}+\frac{v_{\perp}^2}{\kappa \theta_{\perp}^2}\right)^{-\kappa-1},
\end{equation}
with normalization
\begin{equation}
N_{\mathrm{SKD}}=\frac{1}{\pi^{3 / 2} \theta_{\|} \theta_{\perp}^2} \frac{\Gamma(\kappa+1)}{\kappa^{3 / 2} \Gamma(\kappa-1 / 2)} ,
\end{equation}
where $\Gamma$ denotes the Gamma function. 
The parameter $\kappa$ controls the strength of the suprathermal tail and the distribution approaches a Maxwellian in the limit $\kappa\rightarrow\infty$. The thermal speeds are $\theta_{ \|, \perp} = \sqrt{2k_B T^{\kappa}_{\|, \perp}/m}$, with $T^{\kappa}_{\|, \perp}=\frac{\kappa}{\kappa -3/2} T^{M}_{\|, \perp} >T^{M}_{\|, \perp}$ and $\kappa>3/2$.
A well-known limitation of the SKD is that higher-order velocity moments (including essential macro-parameters such as  temperature, heat flux, and transport coefficients) diverge \citep{Scherer-etal-2017}, 
whereas the observed distributions may exhibit stronger tails corresponding to lower values of $\kappa$ \citep{Stverak-etal-2008, Gloeckler-etal-2012, Wilson-etal-2019a, Wilson-etal-2019b}.

To overcome these shortcomings, the regularized $\kappa$ distribution (RKD) has been introduced \citep[][see also \citet{Lazar_Fichtner_Springer2021}]{Scherer-etal-2017}, modifying the SKD by introducing an exponential cut-off that guarantees the convergence of all velocity moments.
\begin{equation}\label{regularisedkappaEQ}
\begin{aligned}
f_{\mathrm{RKD}}\left(v_{\|}, v_{\perp}, \kappa, \alpha \right)= &  N_{\mathrm{RKD}}\left(1+\frac{v_{\|}^2}{\kappa \Theta_{\|}^2}+\frac{v_{\perp}^2}{\kappa \Theta_{\perp}^2}\right)^{-\kappa-1} 
 \exp \left(-\frac{\alpha^2 v_{\|}^2}{\Theta_{\|}^2}-\frac{\alpha^2 v_{\perp}^2}{\Theta_{\perp}^2}\right).
\end{aligned}
\end{equation}
In this case the normalization constant is \citep{Scherer-etal-2019}
\begin{equation}
N_{\mathrm{RKD}}=\frac{1}{\pi^{3 / 2} \Theta_{\|} \Theta_{\perp}^2 W},
\end{equation}
with
\begin{equation}
W = U\left(\frac{3}{2}, \frac{3-2\kappa}{2},\alpha^2 \kappa \right),
\end{equation}
where $U$ denotes the Tricomi function \citep{Oldham2009}. 
The parameter $\alpha>0$ controls the exponential cut-off and is independent of $\kappa$. The 'thermal' speeds are defined as $\Theta_{\|, \perp}$ and differ from those defined in the SKD case.
The RKD retains the desirable suprathermal behavior of $\kappa$ distributions while ensuring that all velocity moments remain finite. 
Consequently, macroscopic plasma quantities are always consistently defined through velocity moments. 
The SKD and Maxwellian distributions are recovered as limiting cases for $\alpha\rightarrow0$ and $\kappa\rightarrow\infty$, respectively, \textcolor{ForestGreen}{see \cite{Scherer-etal-2017, Schroder2025PhPl}}.
Applications of the RKD include the macroscopic description of suprathermal populations, modelling of anisotropic distributions measured in situ in the solar wind, and the evaluation of transport properties in non-equilibrium plasmas \citep{Lazar-etal-2020,Scherer-etal-2021,Husidic-etal-2022,Hau-etal-2023}. 
A more detailed discussion of the RKD and its properties and applications  can be found in \cite{Schroder2025PhPl, Schroder2025ApJ}.

In the present study we extend the investigation of RKD plasmas within linear wave theory by considering arbitrary propagation angles with respect to the magnetic field. 
Earlier studies were largely restricted to parallel EM modes \citep{Husidic_2020, Schroder2025PhPl, Schroder2025ApJ}. 
Here we include oblique propagation and focus on the electron and proton firehose instabilities (EFHI and PFHI, respectively).

Parallel firehose modes are oscillatory or periodic with finite real frequency ($\omega_r\neq0$), whereas oblique propagation also allows for purely non-propagating, aperiodic modes ($\omega_r=0$). 
Linear theory predicts that the fastest-growing electron firehose modes occur at oblique angles and are typically aperiodic \citep{Yoon-etal-1993, Li-Habbal-2000, Shaaban-etal-2019}. 
Hybrid simulations support these predictions \citep{Gary-Nishimura-2003, Lopez-etal-2019}. 
For protons, aperiodic firehose modes can likewise dominate near instability thresholds when mixing with anisotropic electrons \citep{Lopez_etal_2022ApJ}, which is consistent with solar wind observations \citep{Kasper-etal-2002, Bale-etal-2009}. With isotropic electrons, periodic proton firehose modes are expected to dominate \citep{HellingerMatsumoto_JGR_2000}.

While dispersion relations for SKD plasmas can be solved analytically or numerically, e.g., via the DIS-K solver \citep{Lopez_etal_2021_JPP,Lopez2021}, a full (semi-)analytical solvable dispersion tensor and plasma dispersion functions for anisotropic RKD plasmas for oblique propagation has not yet been derived. 
However, such systems can be investigated numerically using the Arbitrary Linear Plasma Solver (ALPS) \citep{Verscharen_2018}. 
This solver determines, for arbitrary propagation angles, the linear Vlasov-Maxwell dispersion relation directly from tabulated background distribution functions and does not require explicit further analytical expressions.

\subsection{Linear dispersion theory}
First, we assume that the plasma fluctuations, specifically those of the electric and magnetic fields (\(\boldsymbol{E}\) and \(\boldsymbol{B}\) respectively), are small enough to justify the application of linear theory. In order to investigate plasma instabilities, we solve the kinetic dispersion relation to obtain the complex frequency
\begin{equation}
    \omega(\boldsymbol{k}) = \omega_r(\boldsymbol{k}) + i\gamma(\boldsymbol{k}),
\end{equation}
where \(\boldsymbol{k}\) is the wave vector corresponding to plane-wave-like fluctuations, and \(\omega_r\) denote the real and \(\gamma\) the imaginary part of the frequency.

To obtain this solution, we use the linearized Vlasov equation and the linearised set of Maxwell's equations, providing an expression for the plasma susceptibilities \(\boldsymbol{\chi}_j\) for the \(j\)-th species (see Appendix \ref{app_disper}). These susceptibilities are then related to the plasma's dielectric tensor \(\boldsymbol{\epsilon}\) through
\begin{equation}
   \boldsymbol{\epsilon} =  \mathbbm{1} + \sum_j \boldsymbol{\chi}_j.
\end{equation}

From that, we can obtain the wave equation 
\begin{equation} \label{wave_eq}
(\boldsymbol{k}c/\omega) \times [(\boldsymbol{k}c/\omega) \times \boldsymbol{E}] + \boldsymbol{\varepsilon} \cdot \boldsymbol{E} \equiv \mathcal{D} \cdot \boldsymbol{E} = 0
\end{equation}
where $c$ is the speed of light. Solving \(\text{det}\ \mathcal{D} = 0\) with $\boldsymbol{E}\neq 0$ provides non-trivial solutions.
To obtain the latter, the ALPS code \citep{Verscharen_2018} will be used, which offers a robust numerical approach, providing solutions for arbitrary linear plasma setups. ALPS has proven (and further proven with this study) to be a valuable tool for studying a wide range of plasma instabilities. For further details on its implementation and capabilities, refer to \citet{Verscharen_2018, Klein_Verscharen_2025PhPl}.

The structure of the paper is as follows. 
We begin with a brief description of our setup in section 2. 
In the first part of this section, the ALPS implementation is validated against a series of previous results, in particular, those obtained for oblique EFHI modes in Maxwellian plasmas \citep{Li-Habbal-2000, Maneva_2016APJ}, and in SKD plasmas \citep{ Shaaban-etal-2019}, and nevertheless for oblique PFHI in Maxwellian plasmas  \citep{HellingerMatsumoto_JGR_2000,Maneva_2016APJ}.
The new RKD results are presented and discussed in the third section. All findings are summarized in the concluding section 4. 

\section{Model setup and validation results}
We focus on the oblique electron and proton firehose instabilities (OEFHI, OPFHI) in a two-component plasma comprising protons ($p$) and electrons ($e$) with one population each. 
The plasma frequency of species $j$ is defined as $\omega_{p,j} = \sqrt{4 \pi n_j q_j^2 / m_j}$, and the corresponding nonrelativistic gyrofrequency as $\Omega_j = q_j B_0 / (m_j c)$, where $q_j$, and $n_j$ denote the electric charge and number density of species $j$, protons ($j = p$) and electrons ($j = e$), respectively. The parallel plasma beta of each species $j$ is $\beta_j = 8 \pi n_j k_B T_{j, \parallel} /B_0^2$, with the magnitude of the background magnetic field $B_0$ and the normalized number density set to $n_e/n_0=n_p/n_0=1.0$. The temperature anisotropy may be defined as $A_j=T_{j, \perp}/T_{j,\|}$. We use a proton-to-electron mass ratio of $m_p/m_e = 1836$ and assume the electron plasma-to-gyrofrequency
ratio to be $\omega_{p,e}/|\Omega_e|=1000$.
By allowing for arbitrary propagation directions, we extend the study by \citet{Husidic_2020}, which was limited to electromagnetic modes parallel to the magnetic field and by \citet{Shaaban-etal-2019}, which was limited to oblique modes of SKDs.
The present investigation explores the effects of RKD distributed plasmas with an isotropic cut-off, $\alpha_{\perp}=\alpha_{\|}\equiv\alpha$ on the dispersion relation for oblique propagation, comparing them with the results obtained for Maxwellian distributions and SKDs. Aperiodic branches with vanishing real frequency are observed in the oblique cases.

\begin{figure}\centering

    \includegraphics[width=0.7\textwidth]{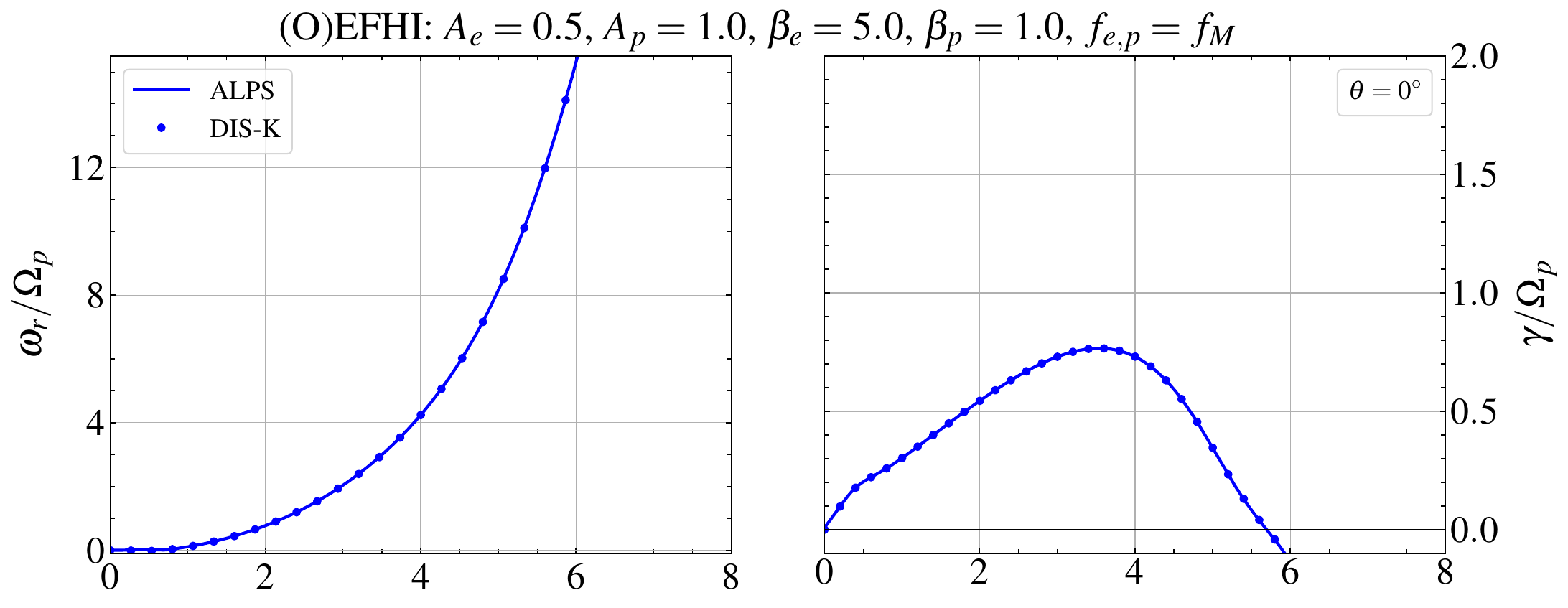}
    \vspace{0.45cm}

    \includegraphics[width=0.7\textwidth]{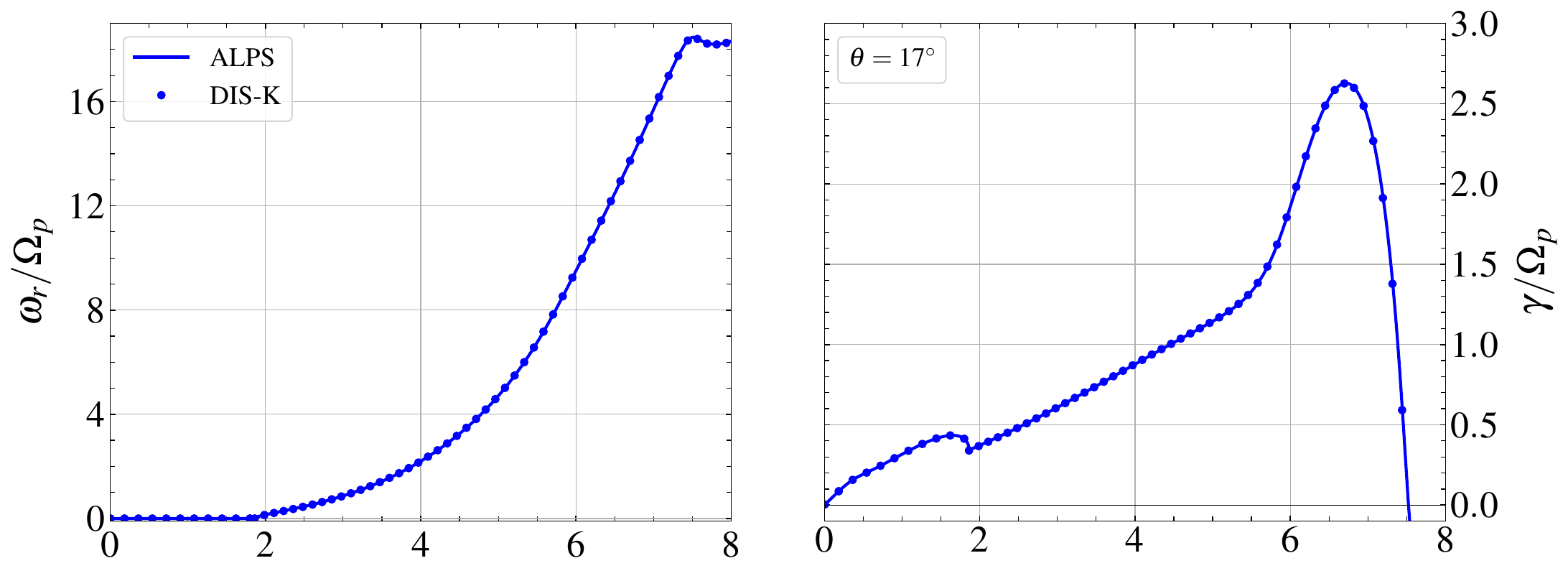}
    \vspace{0.45cm}

    \includegraphics[width=0.7\textwidth]{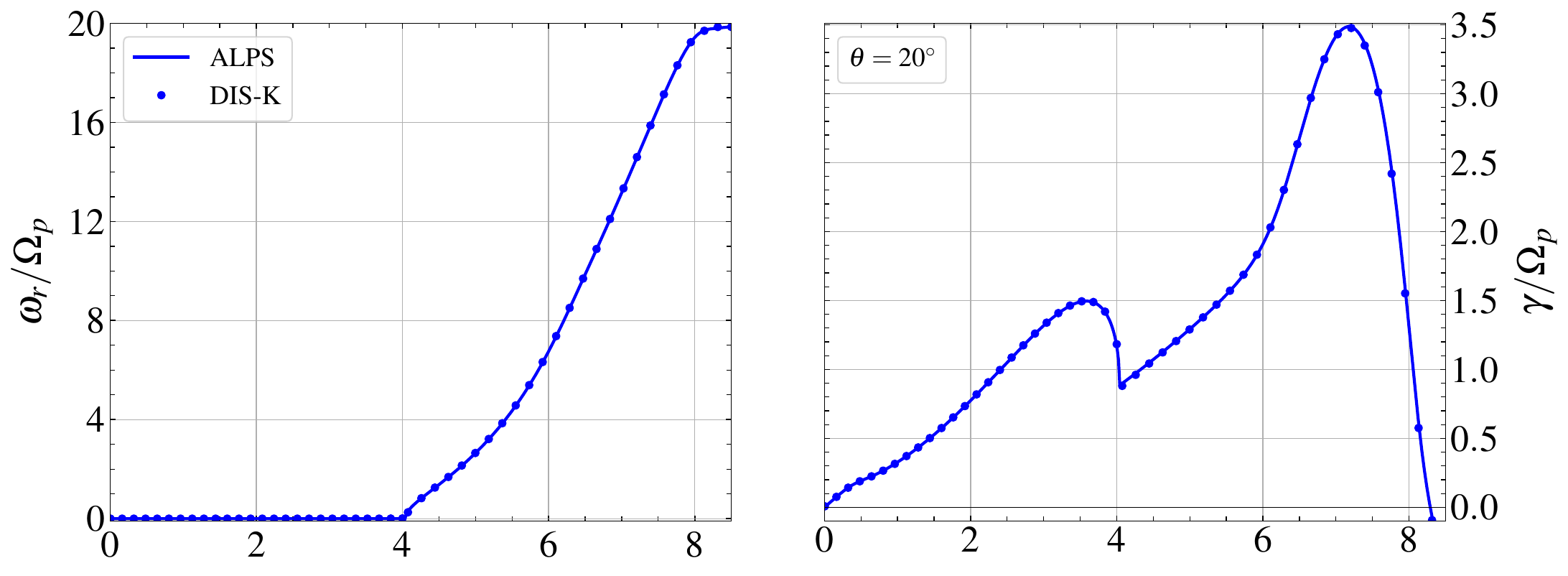}
    \vspace{0.45cm}

    \includegraphics[width=0.7\textwidth]{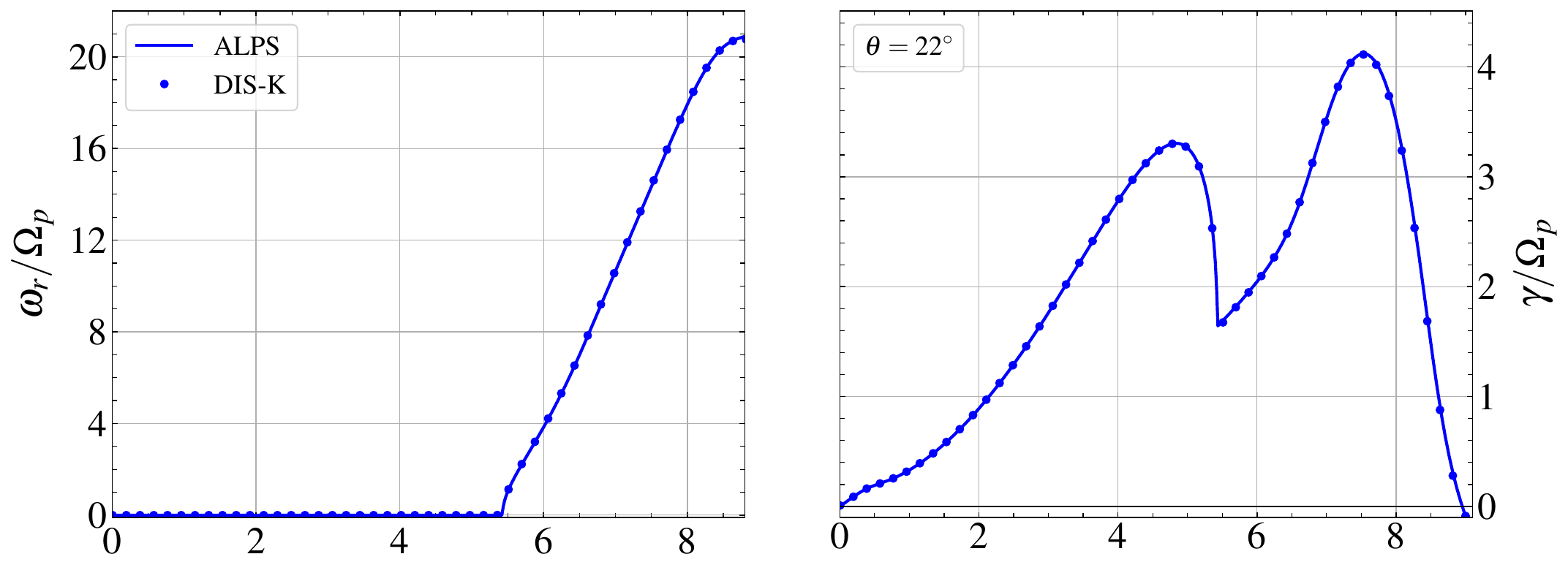}
    \vspace{0.45cm}

    \includegraphics[width=0.7\textwidth]{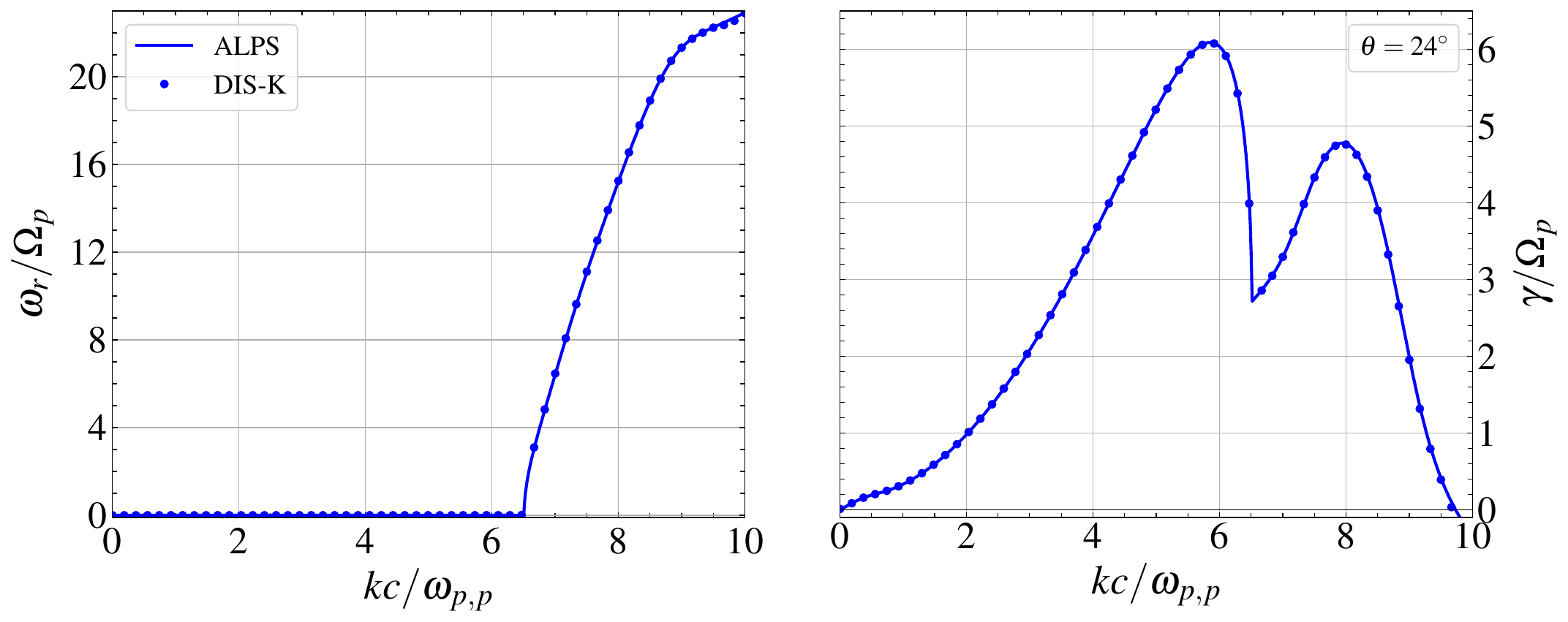}
    \caption{\textit{Normalized wave frequencies (left) and normalized growth rates (right) as functions of the normalized wave number for OEFHI at different propagation angles ($0^\circ$, $17^\circ$, $20^\circ$, $22^\circ$, and $24^\circ$, from top to bottom), obtained with ALPS (solid lines) and DIS-K (dots) for the Maxwellian case.}}
    \label{fig_OEFHI_maxwell}
\end{figure}

\subsection{Validation. OEFHI with Maxwellian and SKD }

To begin with, the ALPS setup is validated against previous results for EFHI driven by bi-Maxwellian electrons \citep{Li-Habbal-2000}, which here are reproduced using DIS-K, imposing the same plasma parameters.
For the electrons, $\beta_{e}=5$ and $T_{e,\perp}/T_{e,\parallel}=0.5$, while the protons are described by an isotropic Maxwellian with $\beta_{p}=1$.
Figure~\ref{fig_OEFHI_maxwell} presents the (O)EFHI dispersion curves for different propagation angles ($0^\circ$, $17^\circ$, $20^\circ$, $22^\circ$, and $24^\circ$, from top to bottom), obtained with ALPS (solid lines) and DIS-K (dots). 
Both the wave frequency (left) and growth rate (right) are shown as functions of the wave number $k=\sqrt{k^2_{\|} + k^2_{\perp}}$.
Overall, the agreement between the two solvers is very good across all propagation angles (less than $0.2\%$ deviation for the maximum growth rate $\gamma_m$ and the corresponding wave number $k_m$). Note that these results were also replicated by NHDS (\cite{Verscharen_2018RNAAS}) and PLUME (\cite{Klein_etal_2025RNAAS}), agreeing with DIS-K and ALPS. 

For parallel propagation, the agreement between Fig.~1a in \citet{Li-Habbal-2000} and our results is excellent.
However, for oblique propagation, noticeable differences arise, as evidenced by a comparison of the unstable solutions in our Figure~\ref{fig_OEFHI_maxwell} and Fig.~1 in
\citet{Li-Habbal-2000}. 
The present results can be regarded as an updated version, and
a detailed comparison between the results reported by \citet{Li-Habbal-2000} and the present ALPS/DIS-K solutions is summarized in Table~\ref{tab:LiHabbal_comparison}.
For the periodic branch, the largest deviations are found at finite propagation angles, where the present solutions exhibit systematically higher growth rates and larger ranges of unstable wave-numbers.
Differences in the aperiodic branch are generally smaller, although with some noticeable deviations in the transition region between periodic and aperiodic modes, see the results for $\theta = 22^\circ$ and $\theta = 24^\circ$.
At $\theta = 22^\circ$, in the present ALPS/DIS-K solutions the periodic peak remains slightly dominant with $\gamma_m$ approximately $1.25$ times larger than the aperiodic peak, in contrast to the dominant aperiodic branch (with $\gamma_m$ approximately $1.5$ times larger than the periodic one) reported by \citet{Li-Habbal-2000}.

More significant differences are found at $\theta = 24^\circ$. While \citet{Li-Habbal-2000} show a strongly dominant aperiodic branch with $\gamma_m$ about $6.7$ times larger than the periodic branch with a non-developed peak,
the present solutions instead exhibit two distinguishable peaks, with only a moderate dominance of the aperiodic one (about $1.27$ times larger than the periodic peak). 
In addition, the periodic branch obtained here remains unstable over a wider wave-number interval with a markedly higher and broader peak.

\begin{figure} 
\centering
    \includegraphics[width=0.75\textwidth]{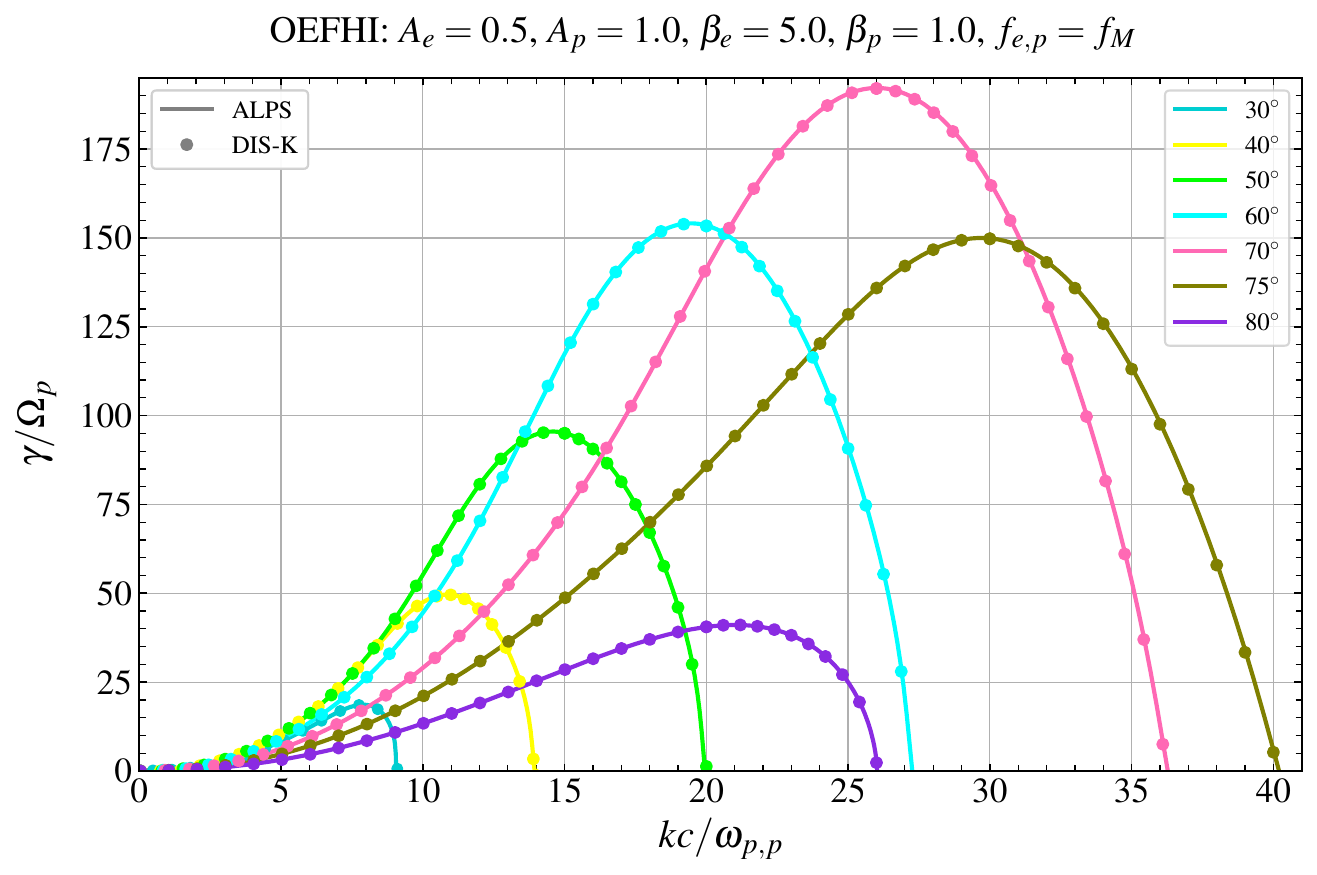}
    \caption{\textit{Normalized growth rates as functions of the normalized wave number for aperiodic OEFHI at different propagation angles ($30^\circ$, $40^\circ$, $50^\circ$, $60^\circ$, $70^\circ$, $75^\circ$, and $80^\circ$), obtained with ALPS (solid lines) and DIS-K (dots) for the Maxwellian case.}}
    \label{fig_OEFHI_maxwell_2}
\end{figure}

\begin{figure}
    \centering
    \includegraphics[width=0.75\textwidth]{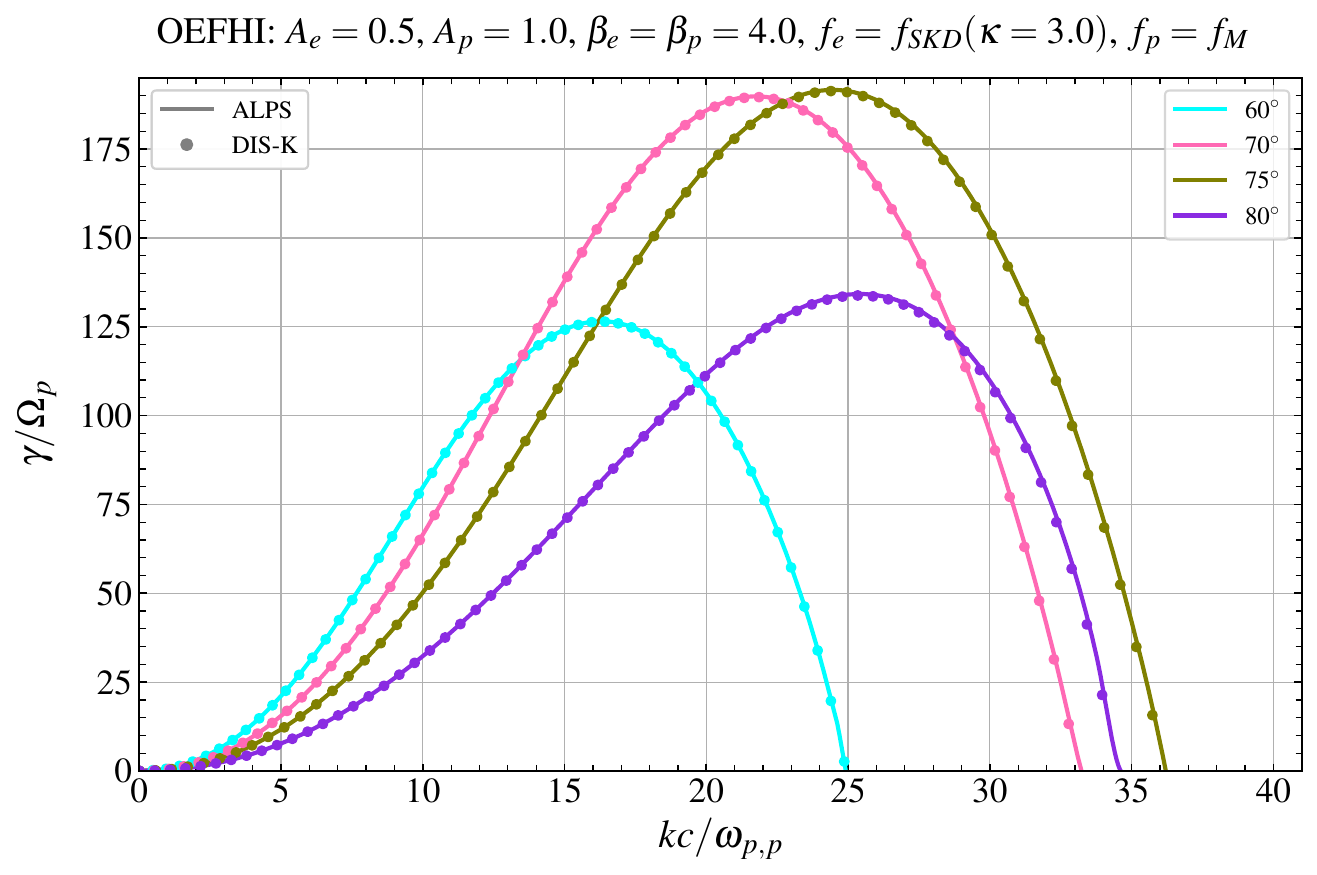}
    \caption{\textit{Normalized growth rates as functions of the normalized wave number, for the oblique (aperiodic) EFHI at propagation angles $60^\circ$, $70^\circ$, $75^\circ$, and $80^\circ$ for SKD with $\kappa =3.0$, obtained with ALPS (solid lines) and DIS-K (dots). }} \label{Oblique_EFHI_skd} 
\end{figure}

Figure~\ref{fig_OEFHI_maxwell_2} shows the growth rates of aperiodic modes at higher propagation angles, 
with ALPS results indicated by solid lines and DIS-K with dots. 
The agreement between both solvers remains very good in all cases (less than $0.1\%$ deviation in terms of $\gamma_m$ and $k_m$). 
The overall solutions are in good qualitative agreement with the results in \citet{Li-Habbal-2000} (see their Fig.~3), including the shift of the growth rates toward higher wave numbers and an increase of $\gamma_m$ up to $\theta=70^\circ$, followed by a decrease at larger angles.
However, we find notable quantitative differences, which are summarized in Table~\ref{tab:LiHabbal_comparison}.
 
The present solutions systematically predict broader unstable wave-number intervals and, in most cases, enhanced maximum growth rates compared to \citet{Li-Habbal-2000}. 
Despite the qualitative agreement, differences are quantitatively noticeable for the periodic branch at finite angles, and for the extent of the unstable range and the maximum growth rates of high-angle aperiodic solutions.

Now we validate ALPS with OEFHI solutions obtained for SKD electrons with DIS-K; see also previous results in \citet{Shaaban-etal-2019}.
Thus, Figure \ref{Oblique_EFHI_skd} shows the comparison of the normalized growth rate between ALPS and DIS-K for the aperiodic OEFHI for different angles ($60^\circ$, $70^\circ$, $75^\circ$, and $80^\circ$) for SKD electrons with $\kappa = 3.0$. 
The agreement between both solvers is very good for all cases (less than $0.4\%$ deviation in terms of $\gamma_m$ and $k_m$).
For bi-Maxwellian electrons, \citet{Li-Habbal-2000} imposed an electron-proton-temperature ratio of $T_{e,\|}/T_{p, \|} = 5$. 
To minimize the effect of the cyclotron resonant protons for the $\kappa$-distributed plasma cases, we use $\beta_e = \beta_p = 4.0$ for the further study of the OEFHI.

\begin{figure}
    \centering
    \includegraphics[width=0.95\textwidth]{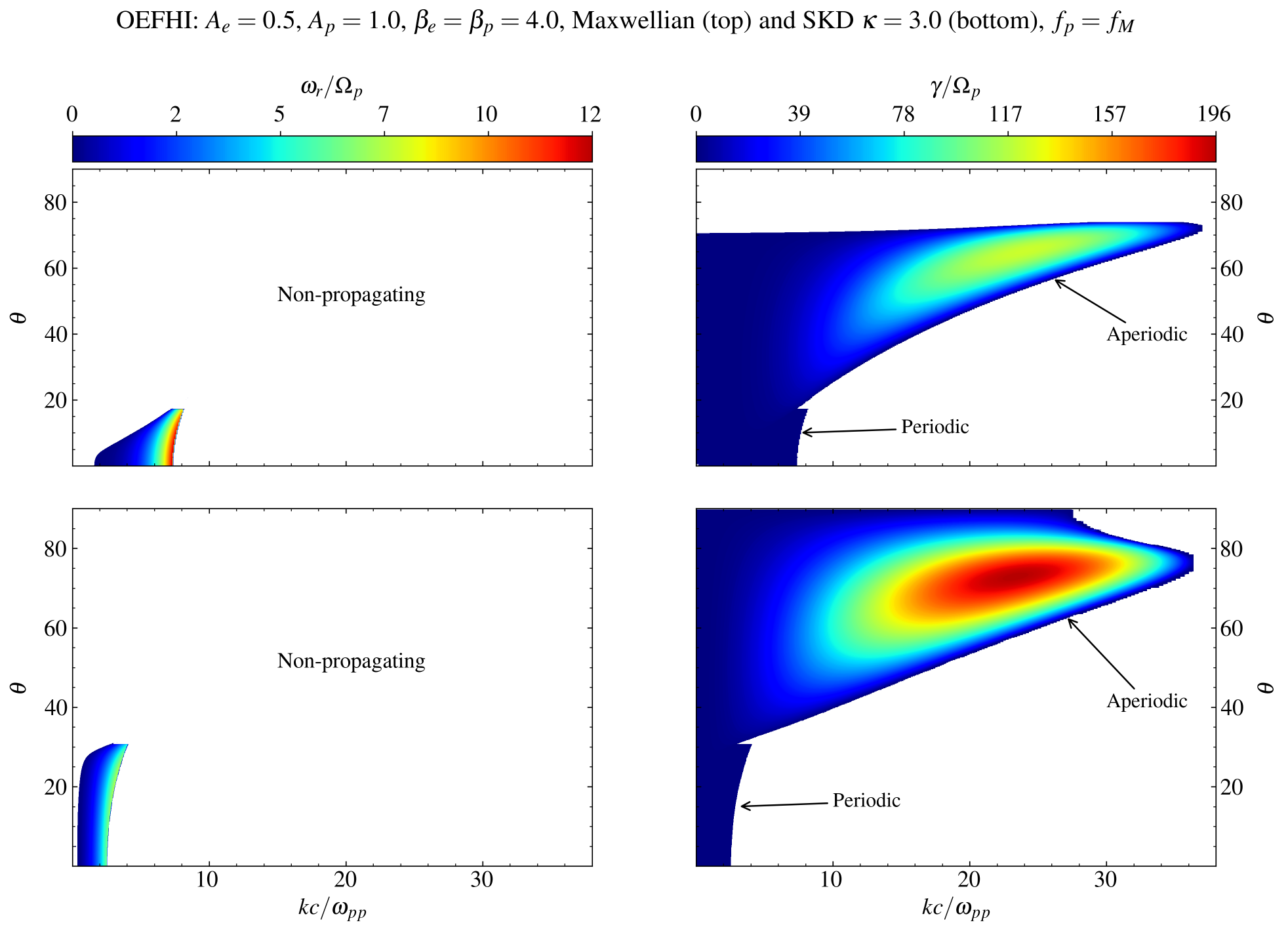}
    \caption{\textit{Contour plot of the OEFHI normalized frequency (left panel) and growth rate (right panel) in the $k$-$\theta$-plane for the Maxwellian case (top) and SKD case with $\kappa=3.0$ (bottom), obtained with DIS-K.}}
    \label{plot_Map_oefhi}
\end{figure}
\vspace{0.35cm}

In Figure \ref{plot_Map_oefhi}, contour plots, derived with DIS-K, of the OEFHI normalized frequency (left panel) and growth rate (right panel) in the $k$-$\theta$-plane are shown for the Maxwellian case (top) and for the SKD with $\kappa =3.0$ (bottom). While the Maxwellian case has unstable periodic solutions up to $\theta \approx 18^\circ$ and $kc/\omega_{pp} \approx 8.1$, the SKD case has unstable periodic solutions up to an increased $\theta \approx 31^\circ$ and decreased $kc/\omega_{pp} \approx 4.1$.
The aperiodic mode is unstable for the Maxwellian case up to $\theta \approx 73^\circ$ and $kc/\omega_{pp} \approx 37$. The maximum growth rate is reached at $\theta_m \approx 65.1^\circ$ with $\gamma_m \approx 122.5$ at $k_m c/\omega_{pp}\approx 24.1$. For the SKD case, the aperiodic mode is unstable until $\theta \approx 89^\circ$ and $kc/\omega_{pp} \approx 36.5$. The maximum growth rate is reached at $\theta_m \approx 72.9^\circ$ with $\gamma_m \approx 195.6$ at $k_m c/\omega_{pp}\approx 23.2$. \\
The validation for the OPFHI, inspired by \cite{HellingerMatsumoto_JGR_2000}, can be found in Appendix \ref{App_Validation_OPFHI}, showing again very good agreement.
These applications to the OEFHI and OPFHI validate both the ALPS and the DIS-K code as reliable tools, because the same results were obtained with rather different numerical algorithms. We note that ALPS is computationally more demanding than DIS-K (or general (semi-)analytic solvers), since it evaluates the dispersion relation through more complex numerical integration schemes over tabulated distribution functions, whereas DIS-K relies on analytical expressions for the SKD plasma dispersion functions. In addition, the findings for the modes in an SKD plasma can now serve as a reference for the RKD ones discussed in the following. 

\section{New results: Oblique Firehose Modes for RKDs}
Imposing plasma parameters identical to those adopted for the validation of the OEFHI and OPFHI cases, the analysis is extended here to RKD plasmas, using ALPS.

\subsection{Oblique electron firehose instability}

\begin{figure}
    \centering

    \includegraphics[width=0.8\textwidth]{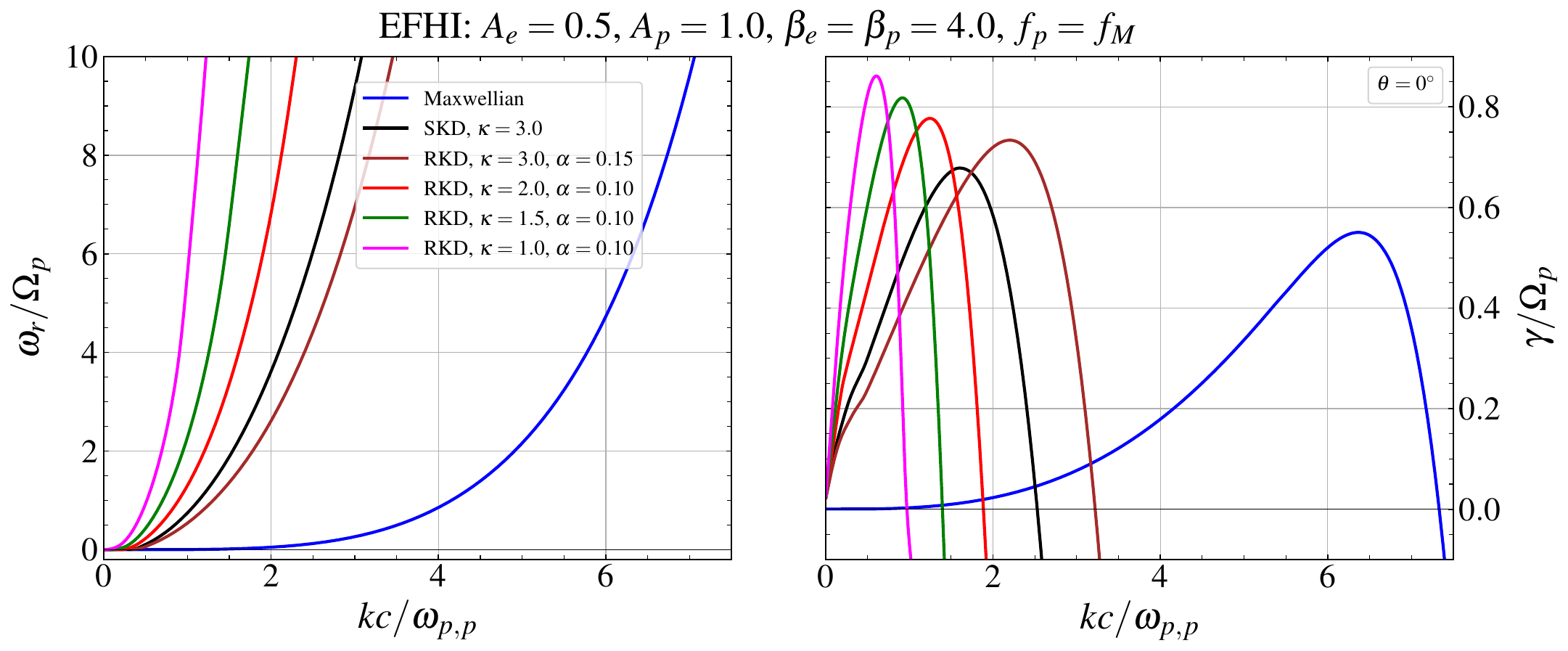}

    \vspace{0.35cm}

    \includegraphics[width=0.8\textwidth]{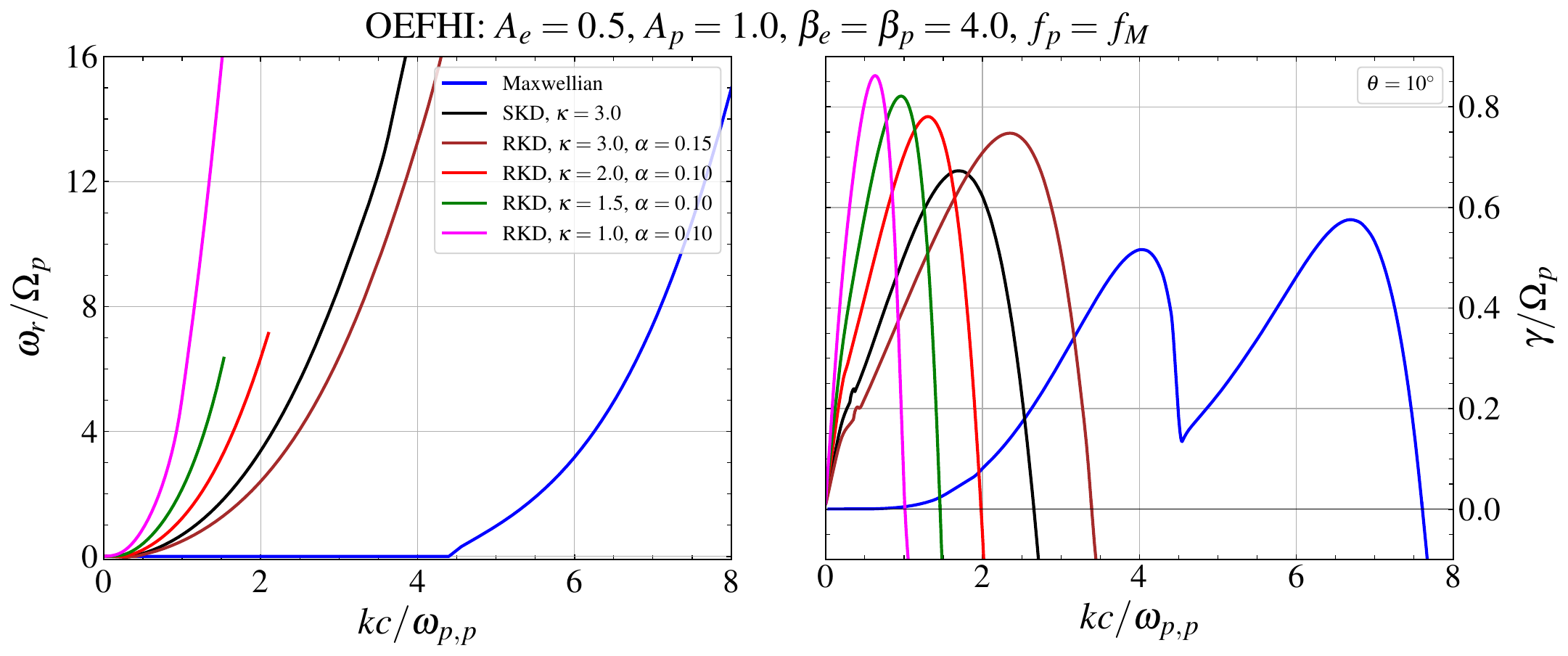}

    \vspace{0.35cm}

    \includegraphics[width=0.8\textwidth]{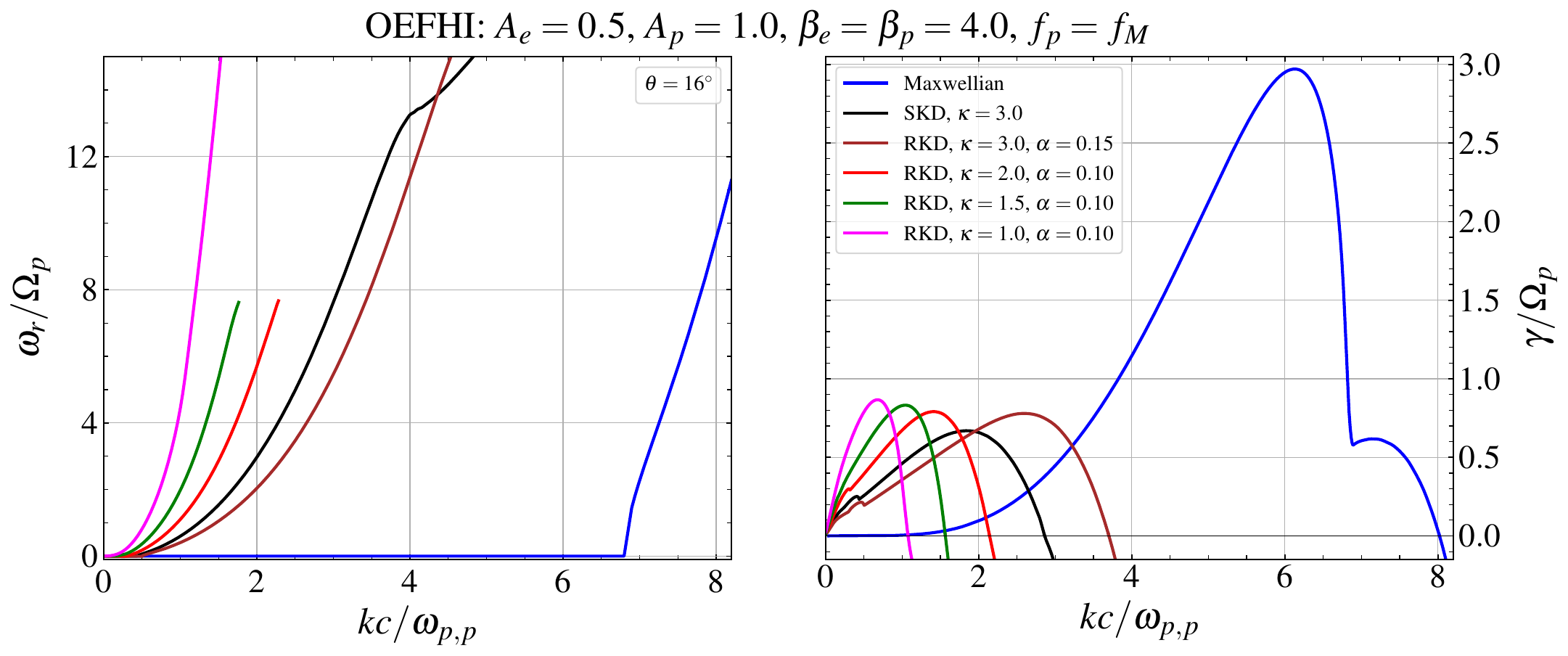}

    \vspace{0.35cm}

    \includegraphics[width=0.8\textwidth]{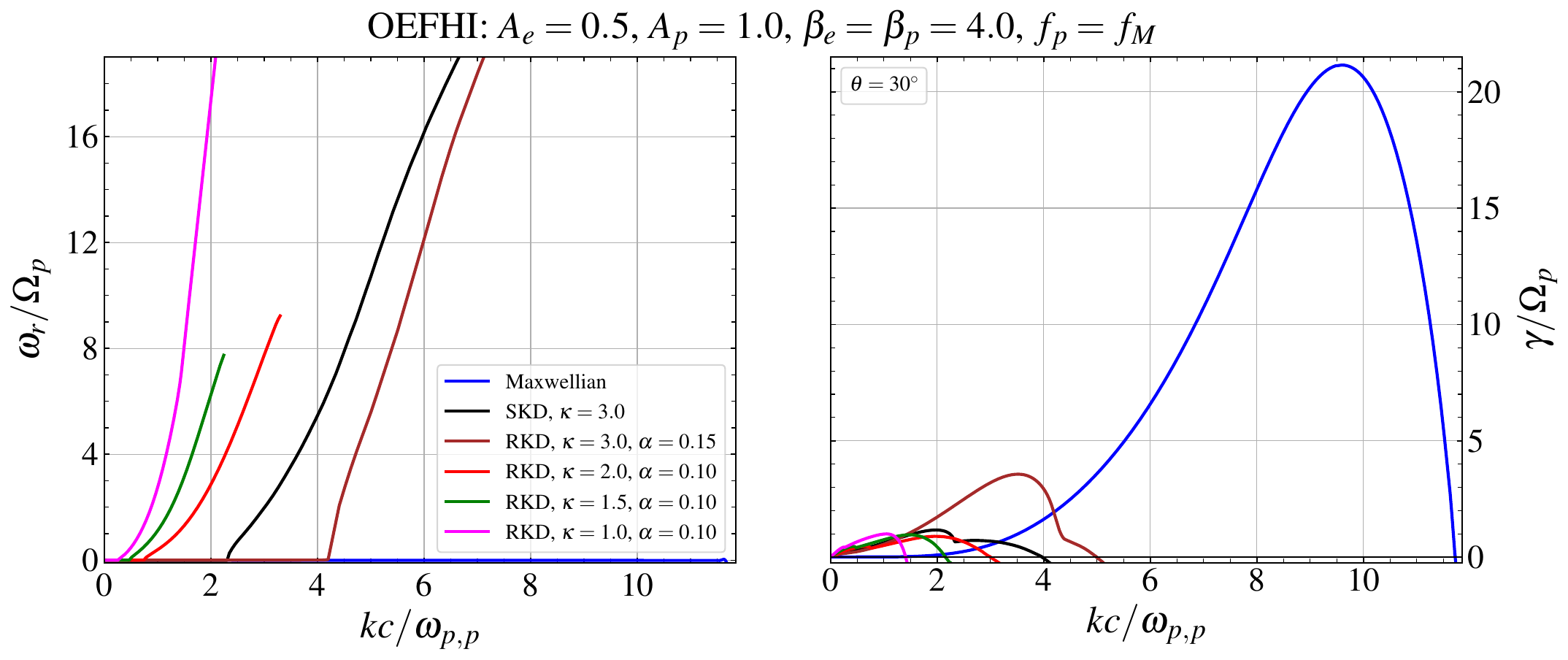}

    \caption{\textit{Normalized wave frequencies (left) and normalized growth rates (right) as functions of the normalized wave number for OEFHI at different propagation angles ($0^\circ$, $10^\circ$, $16^\circ$, and $30^\circ$, from top to bottom), obtained with ALPS for for Maxwellian, SKD, and different RKDs.}}
    \label{Oblique_EFHI_rkd}
\end{figure}

\begin{figure}
    \centering

    \includegraphics[width=0.8\textwidth]{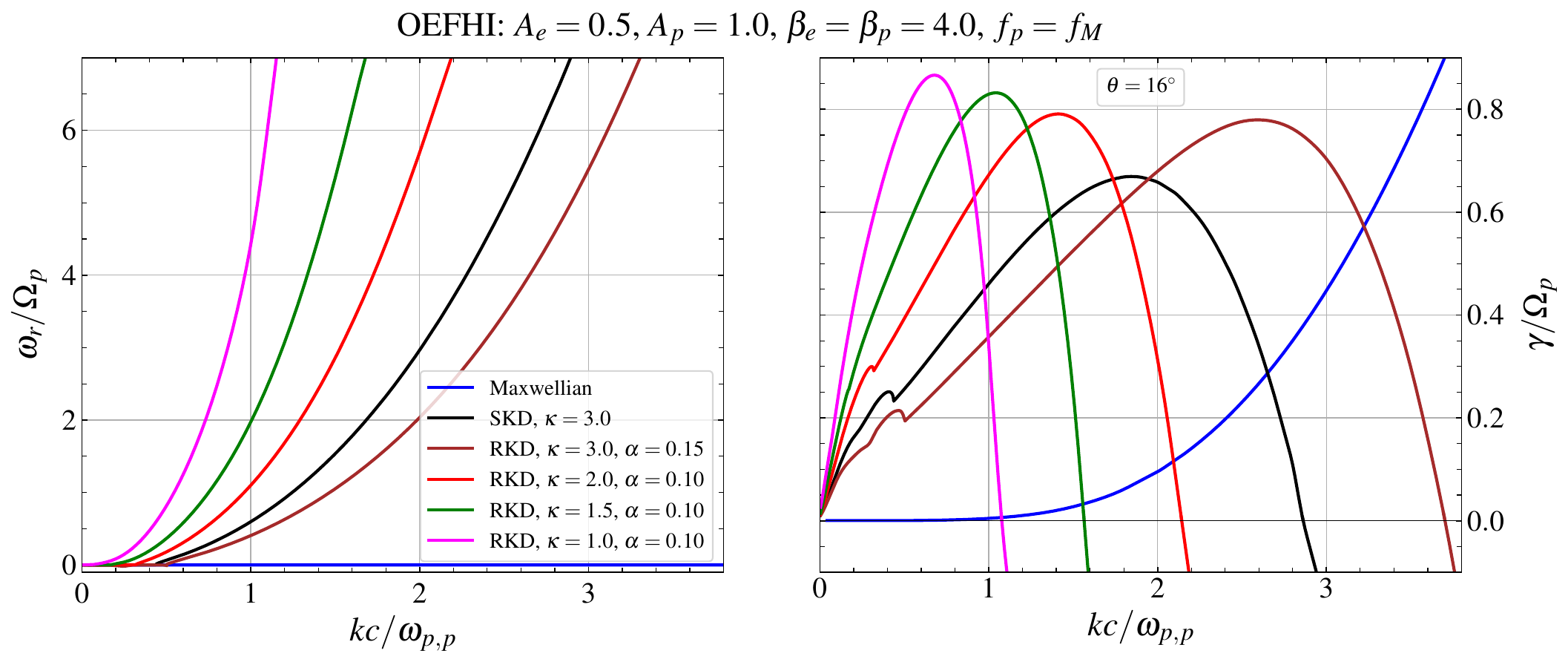}

    \vspace{0.35cm}

    \includegraphics[width=0.8\textwidth]{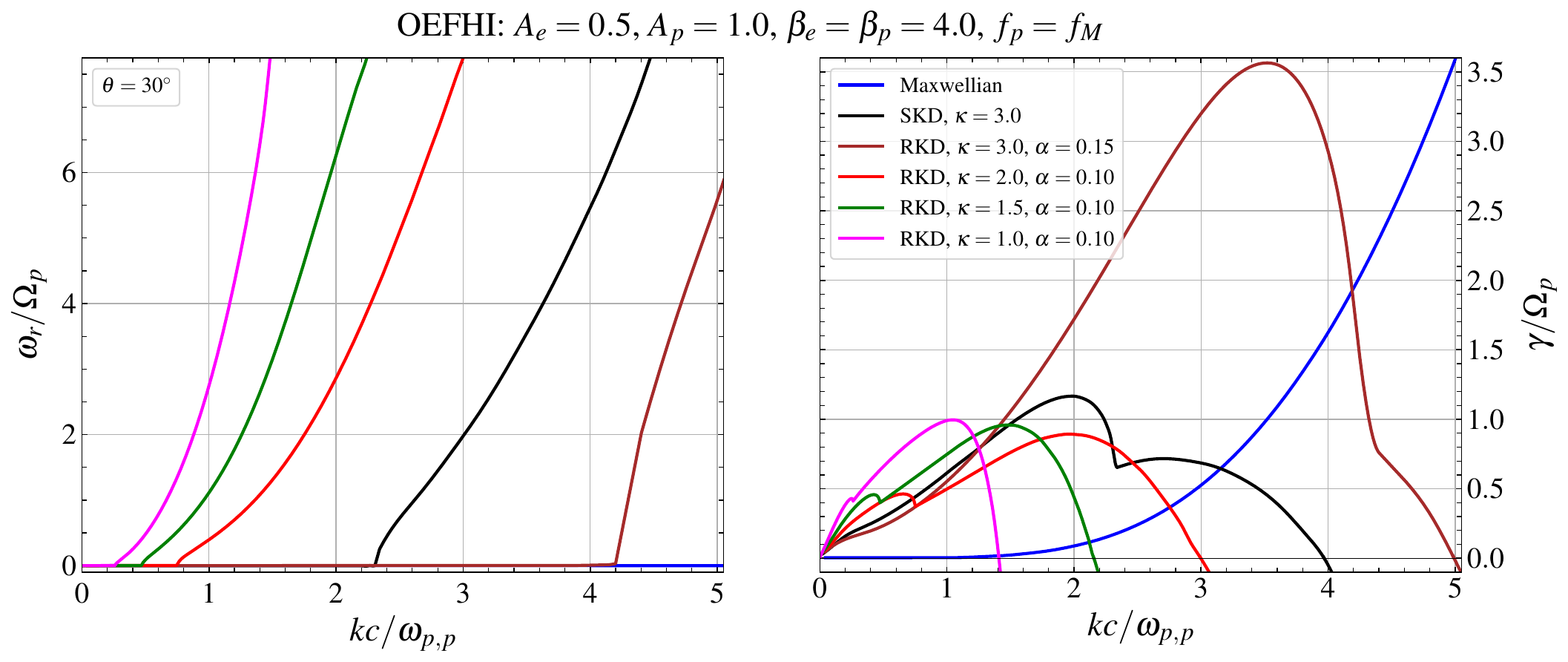}

    \caption{\textit{Normalized wave frequencies (left) and normalized growth rates (right) as functions of the normalized wave number for OEFHI at different propagation angles ($16^\circ$ and $30^\circ$, from top to bottom, same as in Figure \ref{Oblique_EFHI_rkd}, but with more compact plotting scales), obtained with ALPS for Maxwellian, SKD, and different RKDs.}}
    \label{Oblique_EFHI_rkd_zom}
\end{figure}

Figure~\ref{Oblique_EFHI_rkd} shows for the (O)EFHI normalized frequencies in the left panels and normalized growth rates in the right panels for different propagation angles ($0^\circ$, $10^\circ$, $16^\circ$, and $30^\circ$, from top to bottom) and for several velocity distribution functions: Maxwellian (blue), SKD with $\kappa=3.0$ (black),  RKD with $\kappa=3.0$ and $\alpha=0.15$ (brown), RKD with $\kappa=2.0$ and $\alpha=0.10$ (red), RKD with $\kappa=1.5$ and $\alpha=0.1$ (green), and RKD with $\kappa=1.0$ and $\alpha=0.10$ (magenta). Figure~\ref{Oblique_EFHI_rkd_zom} shows the same results for $16^\circ$ and $30^\circ$, but with more compact plotting scales for better visibility of the SKD and RKD solutions. Because all RKDs have different $\kappa$-value, they will be distinguished only by their $\kappa$ values throughout this analysis, for convenience.

The parallel EFHI case, where only periodic modes are present, has been discussed by \citet{Schroder2025PhPl}. The present results are consistent with those findings and can be summarized as follows. Compared to the Maxwellian case, the use of an SKD with $\kappa=3.0$ leads to an increase in the maximum growth rate and a shift toward lower wave numbers, accompanied by a reduction of the unstable wave number range. The real frequency exhibits a similar shift toward lower wave numbers, with enhanced values.
Using an RKD with the same $\kappa=3.0$ value and $\alpha=0.15$, thus reducing the suprathermal tails of the VD, shifts both the real and imaginary solution towards the Maxwellian solution. The maximum growth rate is slightly enhanced, since the moderate cut-off increases the effective anisotropy.
Using RKDs with lower $\kappa$ values and a moderate cut-off ($\alpha=0.1$) maintains the pattern observed for the SKD. As $\kappa$ decreases (while keeping $\alpha$ constant), corresponding to an enhanced population of suprathermal particles, the maximum growth rate increases further and shifts to lower wave numbers, resulting in a more pronounced and narrower peak. For example, while the Maxwellian case yields $\gamma_m/\Omega_p = 0.55$ at $k_m c/\omega_{pp} = 6.36$, the RKD case with $\kappa=1.0$ gives $\gamma_m/\Omega_p = 0.86$ at $k_m c/\omega_{pp} = 0.61$, corresponding to an increase by a factor of $1.56$ and a reduction in $k_m$ by approximately one order of magnitude. Further values regarding the maximum growth rates for the different (O)EFHI cases can be found in Table \ref{tab:max_gamma_oefhi}.

At $\theta = 10^\circ$, the Maxwellian case exhibits a distinct aperiodic peak with $\gamma_m/\Omega_p = 0.516$ at $k_m c/\omega_{pp} = 4.024$, which competes with a periodic branch that remains dominant (we refer to a mode as dominant if its maximum growth rate is higher than the maximum growth rates of all other solutions to the linear dispersion relation), with about $11.6\%$ higher $\gamma_m/\Omega_p = 0.576$ at $k_m c/\omega_{pp} = 6.68$. Compared to the parallel case, the periodic peak shows a very similar maximum growth rate (increase of less than $5\%$), with a slight shift toward higher $k_m$.
For the SKD case, a marginal aperiodic peak appears at low wave numbers, with $\gamma_m/\Omega_p = 0.239$ at $k_m c/\omega_{pp} = 0.363$. The periodic branch clearly dominates, reaching a significantly higher $\gamma_m$, about $182\%$ larger than the aperiodic peak. Relative to the parallel case, the propagating mode remains nearly unchanged in $\gamma_m$, with only a minor shift toward higher $k_m$. The same behavior is seen for the RKD with $\kappa = 3.0$. The aperiodic peak is slightly more pronounced, but is lower in terms of $\gamma_m$ ($\gamma_m/\Omega_p = 0.203$ at $k_m c/\omega_{pp} = 0.408$) compared to the SKD cases.
All three RKD cases with lower $\kappa$-values exhibit only a periodic peak. In comparison to the parallel case, this peak is essentially unchanged in terms of $\gamma_m$ (increase below $0.5\%$), with only a slight shift toward higher $k_m$ ($\lesssim 4\%$). Notably, the periodic peaks of all SKD and RKD cases exceed both the periodic and aperiodic peaks of the Maxwellian case. At certain values of $k$, where $\gamma < 0$, the solutions corresponding to the unstable branch jump to a different branch for the RKDs with $\kappa = 2.0$ and $\kappa = 1.5$. This behaviour is not shown here (nor at other angles), as it is not the focus of the present analysis.
 
At $\theta = 16^\circ$, the Maxwellian case is clearly dominated by the aperiodic branch, with $\gamma_m/\Omega_p = 2.971$, about $382\%$ higher than the periodic peak ($\gamma_m/\Omega_p = 0.617$). The latter shows only a slight increase compared to the parallel case and $\theta = 10^\circ$. However, as the aperiodic branch extends to higher wave numbers, the periodic branch becomes effectively truncated.
For the SKD case, the aperiodic mode becomes more pronounced and slightly enhanced compared to $\theta = 10^\circ$, reaching $\gamma_m/\Omega_p = 0.251$ at $k_m c/\omega_{pp} = 0.408$. Nevertheless, it remains significantly weaker than the periodic mode, which attains $\gamma_m/\Omega_p = 0.669$ at $k_m c/\omega_{pp} = 1.844$. As in the previous cases, the periodic branch shows only a modest increase in $\gamma_m$ and a slight shift toward higher $k_m$. The RKD with $\kappa=3.0$ shows the same development with respect to the periodic peak ($\gamma_m/\Omega_p = 0.780$ at $k_m c/\omega_{pp} = 2.594$) and aperiodic peak ($\gamma_m/\Omega_p = 0.214$ at $k_m c/\omega_{pp} = 0.467$.)
The RKD with $\kappa=2.0$ develops a weak aperiodic mode at $k_m c/\omega_{pp} = 0.307$ with $\gamma_m/\Omega_p = 0.300$ (about $19.5\%$ higher than in the SKD case), together with a dominant periodic mode reaching $\gamma_m/\Omega_p = 0.781$, about $160\%$ larger. The periodic branch remains very similar to the corresponding peaks at lower angles. The RKDs with $\kappa=1.5$ and $\kappa=1.0$ exhibit only a periodic mode, with values closely matching those obtained at smaller propagation angles.
Although the periodic modes in the SKD and RKD cases exceed the corresponding periodic peak of the Maxwellian (by a similar factor as for lower angles), the Maxwellian aperiodic mode has the largest growth rate overall. In particular, it exceeds the periodic mode of the RKD with $\kappa=1.0$ by about $245\%$, and is roughly an order of magnitude larger than the aperiodic mode of the RKD with $\kappa=2.0$.        

At $\theta = 30^\circ$, the Maxwellian case exhibits only a non-propagating mode, with a strongly enhanced growth rate reaching $\gamma_m/\Omega_p = 21.15$ at $k_m c/\omega_{pp} = 9.599$, together with an extended range of unstable wavenumbers.
In contrast to lower angles, the SKD case shows a dominant aperiodic mode, with $\gamma_m/\Omega_p = 2.705$ at $k_m c/\omega_{pp} = 0.717$, exceeding the periodic mode ($\gamma_m/\Omega_p = 1.167$ at $k_m c/\omega_{pp} = 1.984$) by about $63\%$. Using an RKD with $\kappa=3.0$, the periodic peak is barley developed, due to the competition with the aperiodic mode, which gets significantly enhanced (more than an order of a magnitude compared to $\theta=16^\circ$), with $\gamma_m=3.564$ at $k_m c/\omega_{pp} = 3.520$ exceeding the ones of the other SKD/RKD cases (by roughly $670\%$) and showing a clear trend towards the Maxwellian solutions due to the cut-off of the suprathermal tails. 
For the RKD with $\kappa=2.0$, the non-propagating peak is enhanced, reaching $\gamma_m/\Omega_p = 0.463$, which is about $54\%$ higher than at $\theta = 16^\circ$. Nevertheless, the propagating mode remains dominant, with $\gamma_m/\Omega_p = 0.893$, about $93\%$ larger. The RKD cases with lower $\kappa$ values ($\kappa=1.5$ and $\kappa=1.0$) develop aperiodic modes at this angle, in contrast to the behaviour at lower $\theta$. These aperiodic contributions become less pronounced as $\kappa$ decreases. The maximum growth rates, however, remain similar across the RKD cases, with $\gamma_m/\Omega_p = 0.458$ for $\kappa=1.5$ and $\gamma_m/\Omega_p = 0.430$ for $\kappa=1.0$ and about two times smaller than those of the corresponding periodic ones. The periodic peaks are slightly enhanced and shifted toward higher wave numbers. For example, relative to the parallel case, the RKD $\kappa=1.0$ case shows an increase of about $16\%$ in $\gamma_m$ and $72\%$ in $k_m$. Overall, the maximum growth rate in the Maxwellian case exceeds those of the SKD and RKD cases by more than an order of magnitude.

\begin{figure}
    \centering
    \includegraphics[width=\linewidth]{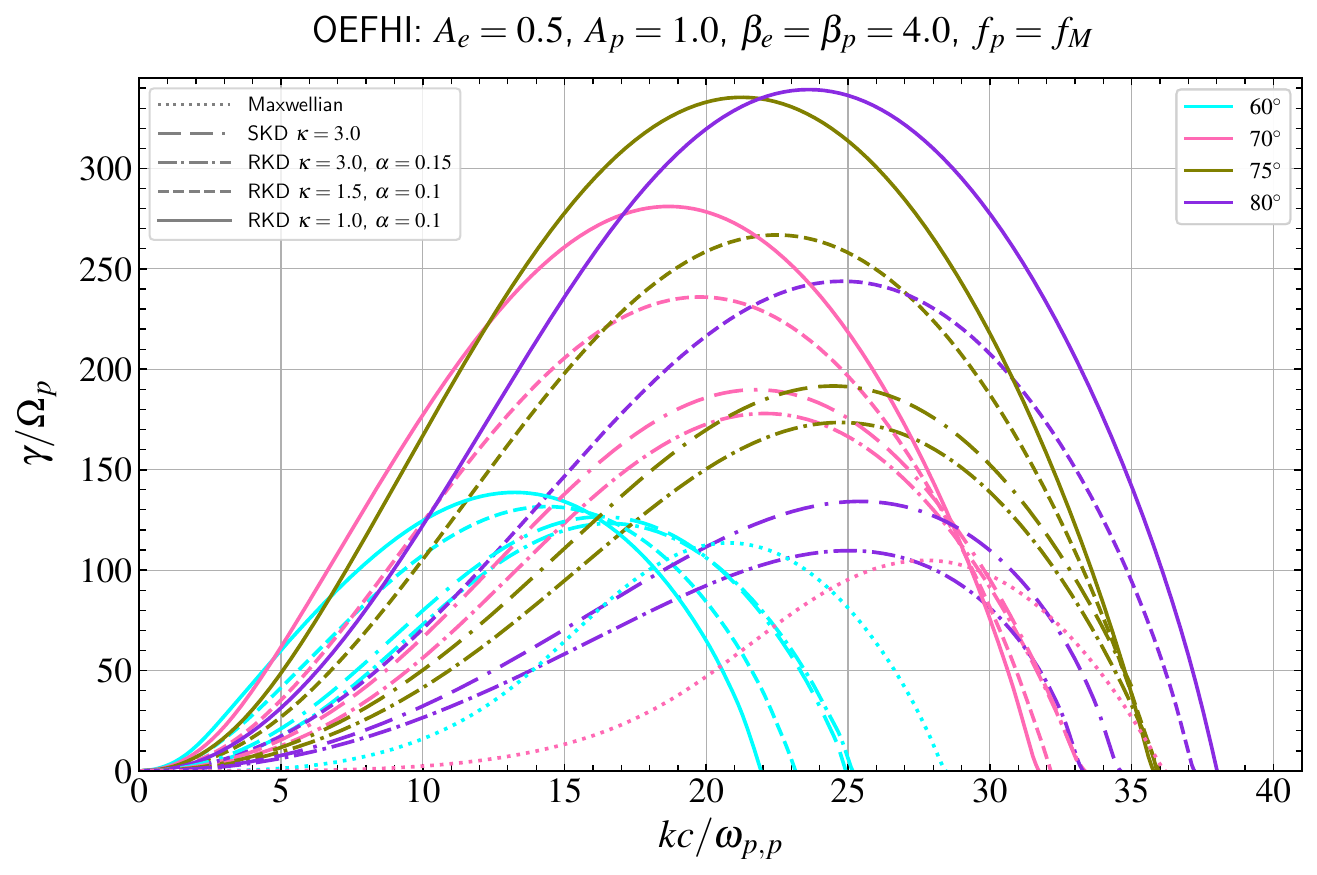}
    \caption{\textit{Normalized growth rates as functions of the normalized wave number, for the (aperiodic) OEFHI at propagation angles $60^\circ$, $70^\circ$, $75^\circ$, and $80^\circ$ for Maxwellian and different RKDs. }} \label{high_oblique_efhi_rkd} 
\end{figure}

Figure~\ref{high_oblique_efhi_rkd} presents the normalized growth rates for higher propagation angles ($60^\circ$, $70^\circ$, $75^\circ$, and $80^\circ$) for the Maxwellian case and for RKDs, obtained with ALPS. At these angles, only non-propagating modes are found, which are in general more than two orders of a magnitude higher then the periodic growth rates at parallel propagation. 
At $\theta = 60^\circ$, the RKD cases exhibit higher growth rates than the Maxwellian case, in contrast to the behaviour at intermediate angles. As $\kappa$ decreases, the maximum growth rate increases and shifts toward lower wave numbers. This trend is consistently observed for all angles considered here. The RKD with $\kappa =3.0$ shows always lower maximum growth rates (about $2\%$-$18\%$) and a shift towards the Maxwellian solution compared to the SKD case.
At $\theta = 70^\circ$, the Maxwellian case has already passed its maximum growth rate (see Figure \ref{plot_Map_oefhi}, where the maximum occurs at $\theta \approx 65^\circ$) and begins to decrease. In contrast, all RKD cases show a significant amplification of the growth rate, accompanied by a shift toward higher wave numbers. For example, the RKD case with $\kappa = 1.0$ reaches $\gamma_m/\Omega_p = 282.1$, corresponding to an increase of $103\%$ relative to $\theta = 60^\circ$ ($\gamma_m/\Omega_p = 138.7$), and an increase of $170\%$ compared to the Maxwellian value $\gamma_m/\Omega_p = 104.6$ at $\theta = 70^\circ$. The Maxwellian case, however, spans a broader range of unstable wave numbers.
The RKD with $\kappa =3.0$ reaches its maximum growth rate at $\theta = 70^\circ$ with $\gamma_m =178.0$ 
At $\theta = 75^\circ$, both $\gamma_m$ and $k_m$ increase further for the RKD cases with lower $\kappa$-values, leading to broad and pronounced peaks. The cases with $\kappa = 2.0$ and $\kappa = 1.5$ reach their maximum growth rates near this angle, with $\gamma_m/\Omega_p = 228.2$ and $\gamma_m/\Omega_p = 266.9$, respectively. The Maxwellian case exhibits no unstable solutions at this angle (see Figure \ref{plot_Map_oefhi}, where instability is limited to $\theta \approx 73^\circ$). 
At $\theta = 80^\circ$, the maximum growth rates for $\kappa = 3.0$ and $\kappa = 1.5$ decrease and shift toward higher wave numbers. In contrast, the case with $\kappa = 1.0$ attains its maximum near this angle, producing a broad peak that extends over a wide wave number interval (up to $kc/\omega_{pp} \approx 38$), with $\gamma_m/\Omega_p = 339.2$ at $k_m c/\omega_{pp} = 23.6$, which is an increase of the growth rate of a factor of roughly 400 compared to the corresponding parallel propagation. These results would not be possible to achieve using SKDs for which the condition $\kappa > 1.5$ must be fulfilled.

\subsection{Oblique proton firehose instability}

\begin{figure}
    \centering

    \includegraphics[width=0.95\textwidth]{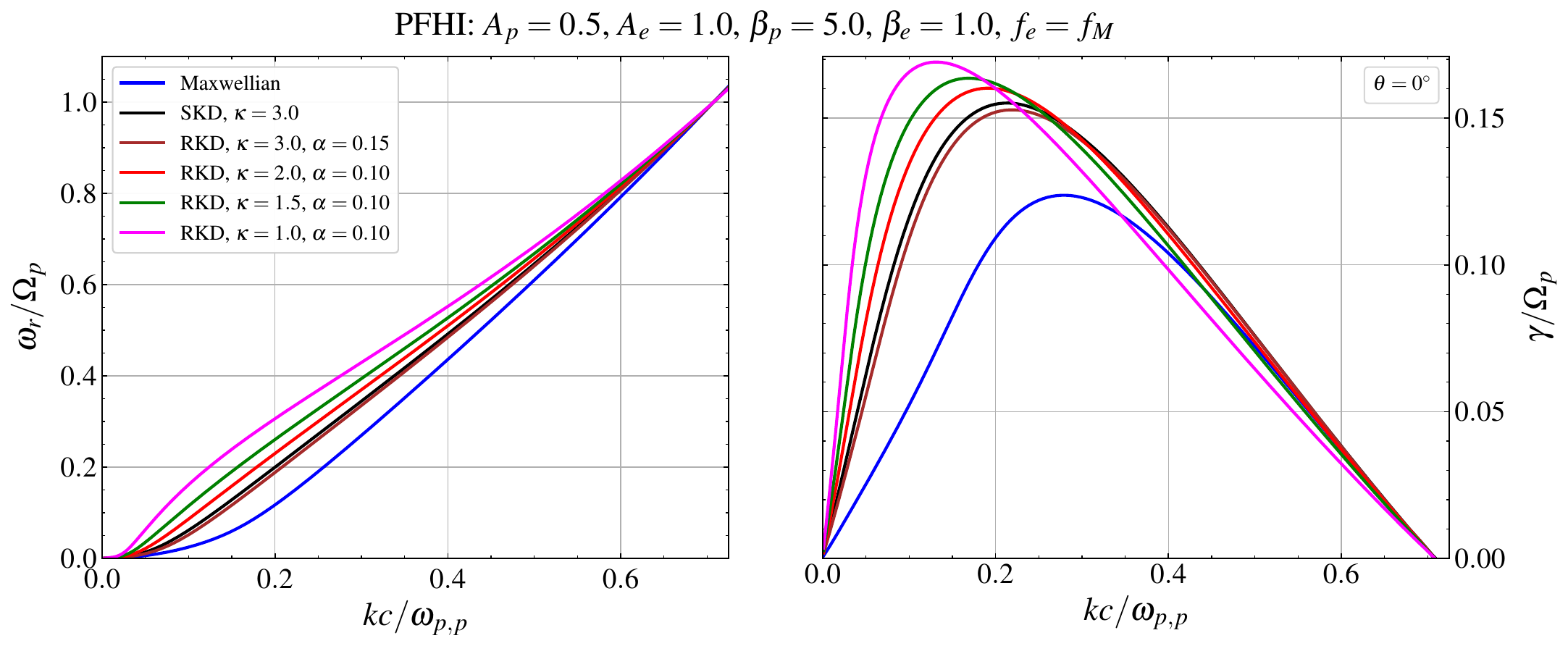}
    \vspace{0.2cm}

    \includegraphics[width=0.95\textwidth]{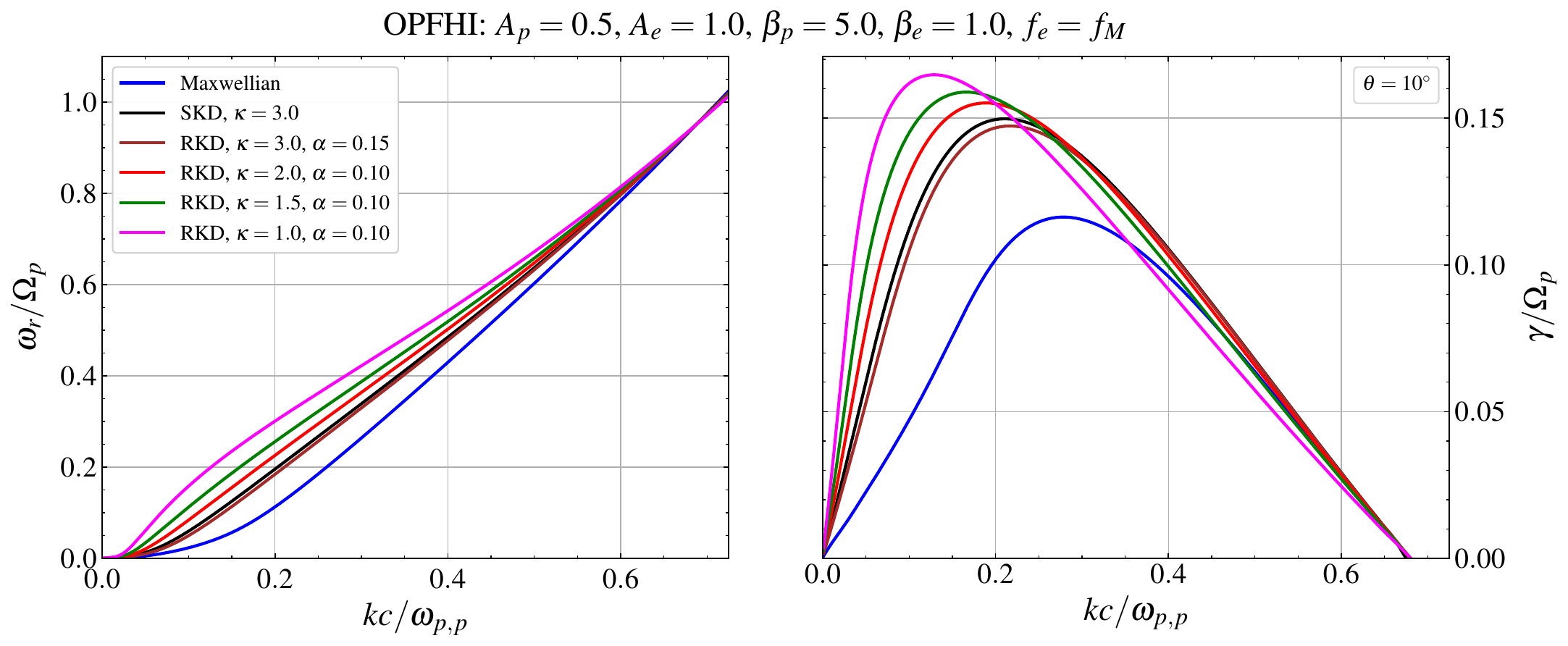}
    \vspace{0.2cm}

    \includegraphics[width=0.95\textwidth]{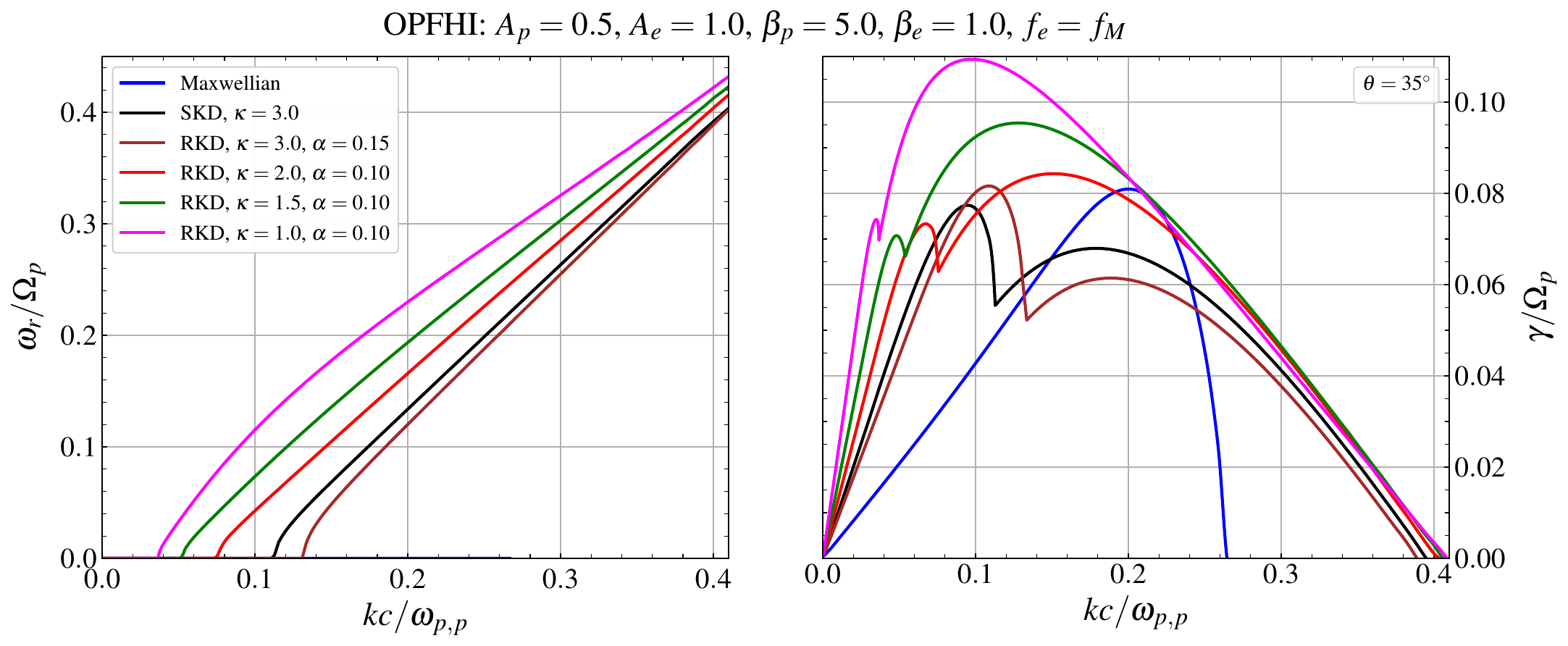}

    \caption{\textit{Normalized wave frequencies (left) and normalized growth rates (right) as functions of the normalized wave number for OPFHI at different propagation angles ($0^\circ$, $10^\circ$, $35^\circ$, from top to bottom), obtained with ALPS for Maxwellian, SKD, and different RKDs.}}
    \label{Oblique_OPFHI_rkd}
\end{figure}

Figure~\ref{Oblique_OPFHI_rkd} shows the (O)PFHI dispersion relations (normalized frequencies in the left panels and normalized growth rates in the right panels) for different propagation angles ($0^\circ$, $10^\circ$, and $35^\circ$, from top to bottom) and for several velocity distribution functions: Maxwellian (blue), SKD with $\kappa=3.0$ (black),  RKD with $\kappa=3.0$ and $\alpha=0.15$ (brown), RKD with $\kappa=2.0$ and $\alpha=0.1$ (red), RKD with $\kappa=1.5$ and $\alpha=0.1$ (green), and RKD with $\kappa=1.0$ and $\alpha=0.1$ (magenta).
 
For the parallel case, where only periodic modes are present, the results are in qualitative agreement with those reported by \citet{Husidic_2020}. They can be summarized as follows. Using an SKD with low $\kappa$ enhances the growth rate and shifts $\gamma_m$ toward lower wave numbers compared to the Maxwellian case. The real frequency is also enhanced at low to intermediate wave numbers. This trend persists when employing RKDs with lower $\kappa$ (at fixed, moderate $\alpha$), corresponding to an increased population of suprathermal particles in the tails. For example, the Maxwellian case yields $\gamma_m/\Omega_p = 0.123$ at $k_m c/\omega_{pp} = 0.279$, whereas the RKD with $\kappa = 1.0$ gives $\gamma_m/\Omega_p = 0.169$ at $k_m c/\omega_{pp} = 0.131$, corresponding to an increase of about $37\%$ in $\gamma_m$ and a reduction of about $113\%$ in $k_m$. Using an RKD with $\kappa=3.0$, the overall growth rate gets slightly reduced (about $2\%$ in terms of $\gamma_m$) and shifted towards the Maxwellian case. The range of unstable wave numbers remains approximately the same for all velocity distributions.

At $\theta = 10^\circ$, only propagating modes are unstable, and only minor differences relative to the parallel case are observed. The maximum growth rate decreases slightly across all cases (by about $3$–$6\%$), while $k_m$ is reduced by $\lesssim 2\%$. Consequently, the unstable wave number range is also slightly reduced.

At $\theta = 35^\circ$, more pronounced differences emerge. In the Maxwellian case, only an unstable aperiodic mode is present, with $\gamma_m/\Omega_p = 0.0809$ at $k_m c/\omega_{pp} = 0.200$, corresponding to reductions of about $34\%$ and $28\%$ relative to the parallel (periodic) case. The unstable wave number range is reduced by more than a factor of two.
In the SKD case, a dominant non-propagating branch appears at low wave numbers, with $\gamma_m/\Omega_p = 0.0774$, which is about $14\%$ higher than for the propagating mode ($\gamma_m/\Omega_p = 0.0679$). The latter is reduced by approximately $50\%$ compared to the parallel case. Both periodic and aperiodic peaks remain below the Maxwellian $\gamma_m$, although the SKD spans a broader unstable wave number range. The RKD case with $\kappa =3.0$ exhibits an $5\%$ increased aperiodic mode ($\gamma_m/\Omega_p = 0.0816$ at higher $k_m c/\omega_{pp} = 0.1088$) and about $10\%$ decreased periodic mode ($\gamma_m/\Omega_p = 0.0614$ at $k_m c/\omega_{pp} = 0.1884$) compared to the SKD case, shifting the behavior towards the Maxwellian solution where only the aperiodic mode is unstable. 
For all three RKD scenarios with lower $\kappa$ values, unstable aperiodic branches are present. As $\kappa$ decreases, these peaks become less pronounced and shift toward lower wave numbers. In all cases, however, the periodic branch remains dominant (e.g. by about $15\%$ for $\kappa=2.0$ and $47\%$ for $\kappa=1.0$). The aperiodic $\gamma_m$ values for the RKDs are smaller than those of the SKD (by about $4$–$9\%$) and the Maxwellian (by about $8$–$13\%$). In contrast, the periodic peaks in all RKD cases exceed those of the SKD (both periodic and aperiodic) and the Maxwellian (aperiodic), although they are reduced and shifted toward lower wave numbers compared to the corresponding parallel RKD cases. For instance, for $\kappa = 1.0$, the periodic peak with $\gamma_m/\Omega_p = 0.1094$ is about $61\%$ higher than the periodic SKD peak, but about $35\%$ lower than in the parallel RKD $\kappa = 1.0$ case. The unstable wave number range in the RKD cases is reduced by approximately a factor of $1.7$ compared to the parallel propagation.

\begin{figure}
    \centering
    \includegraphics[width=\linewidth]{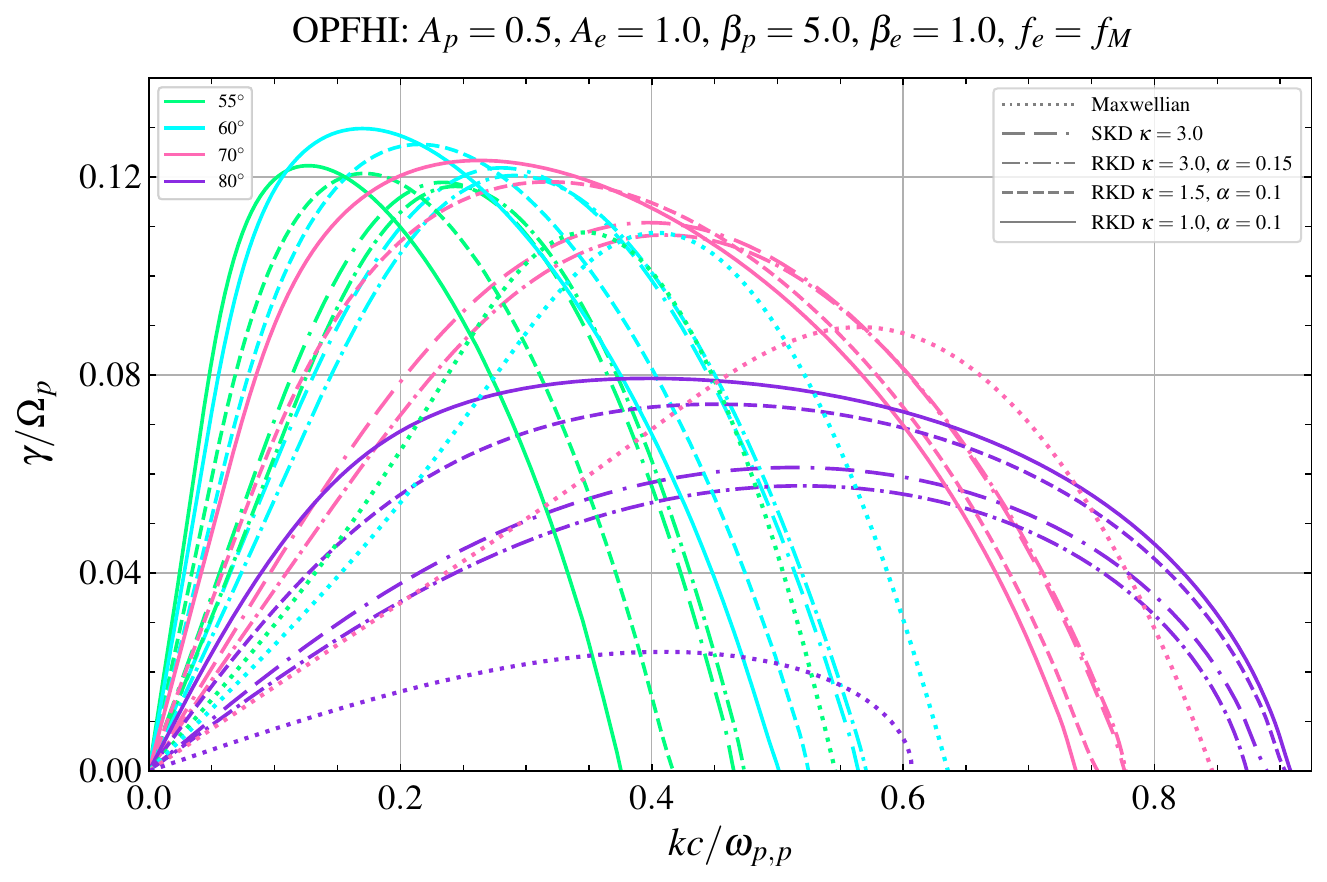}
    \caption{\textit{Normalized growth rates as functions of the normalized wave number, for the (aperiodic) OPFHI at propagation angles $55^\circ$, $60^\circ$, $70^\circ$, and $80^\circ$ for Maxwellian and different RKDs. }} \label{plot_high_OPFHI_rkd} 
\end{figure}

Figure~\ref{plot_high_OPFHI_rkd} shows the normalized growth rates at higher propagation angles ($60^\circ$, $70^\circ$, $75^\circ$, and $80^\circ$) for the Maxwellian and SKD case and several RKDs, computed with ALPS. In this angular range, only unstable non-propagating modes are observed. The maximum growth rates in the RKD cases exceed those of the Maxwellian case at all angles. As $\kappa$ decreases, $\gamma_m$ increases and shifts toward lower $k_m$. For example, at $\theta = 70^\circ$, the RKD case with $\kappa = 1.0$ yields $\gamma_m/\Omega_p = 0.1233$ at $k_m c/\omega_{pp} = 0.2629$, which is about $38\%$ higher and corresponds to a $36\%$ reduction in $k_m$ compared to the Maxwellian case ($\gamma_m/\Omega_p = 0.0897$ at $k_m c/\omega_{pp} = 0.4093$). The RKD with $\kappa=3.0$ shows slightly lower maximum growth rates (up to $6\%$ ) for all cases compared to the SKD, with differences becoming more pronounced at the highest angles, shifting the behavior towards the solution of the Maxwellian VD.

As $\theta$ increases, $\gamma_m$ increases up to a maximum angle, while $k_m$ also increases and the peaks become significantly broader (particularly at $\theta = 80^\circ$), extending over a wider interval of unstable wave numbers. For all velocity distributions, the maximum aperiodic growth rates remain below the corresponding maximum periodic growth rates in the parallel case. For instance, the Maxwellian case reaches its maximum aperiodic growth rate around $\theta = 55^\circ$ with $\gamma_m/\Omega_p = 0.1088$, which is about $12\%$ lower than the parallel value $\gamma_m/\Omega_p = 0.123$.

In all RKD cases, the maximum aperiodic growth rate is attained at slightly larger angles, around $\theta = 60^\circ$. For example, for $\kappa = 1.0$, $\gamma_m/\Omega_p = 0.1298$, which is about $23\%$ lower than the corresponding parallel value $\gamma_m/\Omega_p = 0.169$ at $\theta = 0^\circ$. At $\theta = 80^\circ$, the RKD with $\kappa = 1.0$ exhibits a maximum growth rate that is about $229\%$ higher than in the Maxwellian case, along with an increase of about $50\%$ in the range of unstable wave numbers.

\section{Summary and outlook}
In this work, we investigate the oblique electron and proton firehose instabilities in plasmas with velocity distributions described by Maxwellians, by SKDs, and for the first time by RKDs. The analysis extends previous studies of parallel propagation and provides a systematic comparison of the role of suprathermal particles for oblique wave propagation.

For the OEFHI, the validation against \citet{Li-Habbal-2000}, while showing very good agreement between ALPS and DIS-K for Maxwellian plasmas, reveals systematic quantitative differences for oblique propagation. In particular, the present results consistently show an extension of the unstable wave number range and significantly enhanced growth rates for the periodic branch at intermediate angles. These differences lead to modified relative contributions of periodic and aperiodic branches, such that the significance of these modes can differ from earlier findings. The results presented here can therefore be regarded as an updated and more resolved reference for OEFHI in Maxwellian plasmas.

The comparison between Maxwellian, SKD, and RKD plasmas reveals systematic trends. For parallel propagation, only periodic modes are present. Introducing suprathermal particles via SKDs enhances the growth rates and shifts the maxima toward lower wave numbers. This trend becomes more pronounced for RKDs as $\kappa$ decreases, i.e. with increasing suprathermal particles, leading to sharper and more localized peaks in $k$-space.

With increasing propagation angle, the behaviour becomes more complex due to the emergence of unstable aperiodic modes. For the Maxwellian case, the transition from periodic to aperiodic dominance occurs already at relatively small angles, and at intermediate angles the aperiodic branch can strongly dominate and exceed the growth rate of SKD and RKD. At larger angles, only non-propagating modes remain, with a pronounced maximum growth rate at highly oblique propagation.

In contrast, SKD and RKD plasmas show a '$\theta$-delayed' 
and weaker development of aperiodic modes. For small to intermediate angles, the periodic branch remains dominant in most cases, even when aperiodic contributions are present. As $\kappa$ is reduced, the periodic modes are systematically enhanced and shifted to lower wave numbers, while the aperiodic modes become less pronounced or appear only at larger angles. There are also scenarios in which, at certain intermediate propagation angles, the aperiodic mode of a $\kappa$-distribution attains a higher growth rate than both the periodic and aperiodic modes of a distribution with a lower $\kappa$ value, in contrast to the behaviour observed at smaller and larger angles: At highly oblique propagation, RKDs exhibit significantly enhanced growth rates compared to Maxwellian plasmas, with broader peaks and extended unstable wave number ranges. Notably, the maximum growth rate shifts to larger angles for lower $\kappa$, and the overall growth rates become substantially increased by suprathermal effects. These growth rates would not be achievable using SKDs due to their limitation to $\kappa > 3/2.$ Compared to the periodic modes at parallel propagation, the aperiodic modes at higher angles are about two orders of a magnitude or more higher, making the latter more effective in  regulating the free energy and thus reducing the anisotropy. 

For the OPFHI, a qualitatively similar dependence on the velocity distribution is observed, but with important differences in the angular behaviour. In contrast to the OEFHI, the maximum growth rates are found for parallel propagation for both Maxwellian and SKD plasmas, as already shown in other studies and here also observed for RKDs. As the angle increases, the the growth rates for the periodic solutions decrease and eventually only aperiodic modes remain. The inclusion of suprathermal particles again enhances the growth rates and shifts the maxima toward lower wave numbers. This trend strengthens with decreasing $\kappa$ (for RKDs), leading to systematically higher growth rates than in the Maxwellian case at all angles. However, even at high obliqueness, 
the maximum growth rates of the aperiodic solutions remain below the corresponding values for the parallel (periodic) solutions.

While already shown for Maxwellian and SKDs, the OEFHI for RKDs with low-$\kappa$ transitions to aperiodic dominance at increased intermediate angles and exhibits its highest growth at oblique propagation. The growth rates for the OPFHI (with this set of plasma parameters) remain primarily dominated by periodic modes at parallel propagation, with aperiodic modes becoming relevant only at larger $\theta$ and never exceeding the parallel growth rates.

The comparison between the SKD with $\kappa=3$ and the corresponding RKD with $\kappa=3$ and $\alpha =0.15$ reveals that the cut-off generally shifts the solutions toward the Maxwellian case. For the OEFHI, this effect is particularly pronounced at intermediate propagation angles, where the suppression of the suprathermal tail enhances the relative importance of the aperiodic branch and can substantially increase its growth rate compared to the SKD case. At highly oblique propagation, however, the cut-off reduces the growth rates (of the aperiodic modes) and moves the instability closer to those of the Maxwellian plasma. Similar trends are found for the OPFHI.

The excellent agreement between ALPS and DIS-K demonstrates that ALPS is a reliable and flexible tool for studying oblique instabilities in plasmas with RKDs. This is particularly important since semi-analytical formulations for oblique propagation with RKDs are not yet  available and numerical approaches remain essential for exploring their linear dispersion relations.

At the same time, the generation of contour maps in the $(k,\theta)$-plane, which are valuable for identifying global maxima of the growth rate and for a systematic characterization of oblique instabilities, is computationally more complex with ALPS compared to (semi-)analytic solutions. The development of semi-analytical expressions for oblique RKDs
\citep[analogous to those for isotropic RKDs derived by][]{Gaelzer-etal-2024}, or improved numerical tools for efficiently obtaining these maps, would therefore be beneficial for future investigations.

Overall, the results demonstrate that suprathermal particles, as described by low-$\kappa$ RKDs, have a strong impact on the oblique firehose instabilities, modifying both the angular dependence of the maximum growth rate and the interplay between periodic and aperiodic modes. These effects, while to be expected, become indeed increasingly pronounced with lower $\kappa$, scenarios which are inaccessible using SKDs, reflecting the role of suprathermal particles in enhancing the free energy available to drive kinetic instabilities. In this sense, the RKD adds the ability to actually realize and quantify this regime in a physically consistent way. The value of the present results, therefore, lies not only in the (anticipated) qualitative trend itself and not only in quantifying how the growth rates and unstable wave-number ranges are enhanced once the low $\kappa$ regime becomes accessible, but more in demonstrating that this trend indeed persists for oblique propagation and for both the electron and proton firehose instabilities.

Beyond the specific parameter choices explored here, these results are relevant for the interpretation of temperature-anisotropy constraints observed in the solar wind and other weakly collisional space and astrophysical plasmas, where suprathermal populations are ubiquitous and may be characterized by RKDs with low
$\kappa$ values. The present work illustrates that physically consistent modeling of the low-$\kappa$ regime, made possible by the RKD, is not merely a technical refinement of the SKD but may have direct consequences for how kinetic instabilities are expected to constrain macroscopic plasma parameters in observational and simulation studies.

\section*{Acknowledgements}
We gratefully acknowledge Rodrigo A. López for his ongoing support with DIS-K and for valuable discussions and technical assistance. The ALPS project received support from UCL’s Advanced Research Computing Centre through the Open Source Software Sustainability Funding scheme. This research was supported by the International Space Science Institute (ISSI) in Bern, through ISSI International Team project \#612 (Excitation and Dissipation of Kinetic-Scale Fluctuations in Space Plasmas) led by K.~G.~Klein.
\section*{Funding}
The authors acknowledge support from the Ruhr-University Bochum and the Katholieke Universiteit Leuven. This project was funded by the Deutsche Forschungsgesellschaft (DFG, project FI $706/31$-$1$), the Belgian FWO-Vlaanderen (G$002523$N), and SIDC Data Exploitation (ESA Prodex, No. $4000145223$). D.V. is supported by STFC Consolidated Grant ST/W001004/1.
K.G.K. was supported in part by NASA contract 80NSSC24K0171.
\section*{Declaration of interests}
The authors report no conflict of interest.

\appendix

\section{Dispersion relation in ALPS}\label{app_disper}
ALPS uses the general expression of the plasma's susceptibilities \citep{Verscharen_2018}
\begin{equation}
\boldsymbol{\chi}_j=\frac{\omega_{\mathrm{p}, j}^2}{\omega \Omega_{j}} \int_0^{\infty} 2 \pi p_{\perp} \mathrm{d} p_{\perp} \int_{-\infty}^{+\infty} \mathrm{d} p_{\|}\left[\hat{\boldsymbol{e}}_{\|} \hat{\boldsymbol{e}}_{\|} \frac{\Omega_{j}}{\omega}\left(\frac{1}{p_{\|}} \frac{\partial f_{0 j}}{\partial p_{\|}}-\frac{1}{p_{\perp}} \frac{\partial f_{0 j}}{\partial p_{\perp}}\right) p_{\|}^2\right. 
\left.+\sum_{n=-\infty}^{+\infty} \frac{\Omega_{j} p_{\perp} U}{\omega-k_{\|} v_{\|}-n \Omega_{j}} \boldsymbol{T}_n\right],
\end{equation}
 
 where
 \begin{equation}
U \equiv \frac{\partial f_{0 j}}{\partial p_{\perp}}+\frac{k_{\|}}{\omega}\left(v_{\perp} \frac{\partial f_{0 j}}{\partial p_{\|}}-v_{\|} \frac{\partial f_{0 j}}{\partial p_{\perp}}\right),
\end{equation}

\begin{equation}
\renewcommand{\arraystretch}{2.7}
\boldsymbol{T}_n \equiv
\begin{pmatrix}
\dfrac{n^{2} J_n^{2}}{z^{2}}
& \dfrac{i n J_n J_n'}{z}
& \dfrac{n J_n^{2} p_{\|}}{z p_{\perp}} \\
-\dfrac{i n J_n J_n'}{z}
& \left(J_n'\right)^{2}
& -\dfrac{i J_n J_n' p_{\|}}{p_{\perp}} \\
\dfrac{n J_n^{2} p_{\|}}{z p_{\perp}}
& \dfrac{i J_n J_n' p_{\|}}{p_{\perp}}
& \dfrac{J_n^{2} p_{\|}^{2}}{p_{\perp}^{2}}
\end{pmatrix},
\end{equation}
$z \equiv k_{\perp} v_{\perp} / \Omega_{j}$, and $J_n \equiv J_n(z)$ denotes the $n$ th-order Bessel function. Using the plasma's dielectric tensor as 
\begin{equation}
    \boldsymbol{\epsilon} = \mathbbm{1} + \sum_j \boldsymbol{\chi}_j,
\end{equation}
where $\mathbbm{1}$ is the unity tensor, the wave equation can be written as in equation \ref{wave_eq}.

\section{Validation and comparison of OEFHI solutions}
Table~\ref{tab:LiHabbal_comparison} summarizes the main quantitative differences between the present ALPS/DIS-K solutions and the OEFHI results previously reported by \cite{Li-Habbal-2000} for selected propagation angles, comparing the extent of the unstable wave-number range, the location of the maximum growth rate, and the relative dominance of periodic and aperiodic modes. 

\begin{table*}
\centering
\caption{\textit{Comparison between the OEFHI results reported by \citet{Li-Habbal-2000} and the present ALPS/DIS-K solutions for different propagation angles.}}
\label{tab:LiHabbal_comparison}
\renewcommand{\arraystretch}{1.2}
\resizebox{\textwidth}{!}{
\begin{tabular}{c c c c c c c c}
\hline
\hline
$\theta$ &
Mode &
Source &
$kc/\omega_{pp}$ range &
$k_m c/\omega_{pp}$ &
$\gamma_m/\Omega_p$ &
Difference in $\gamma_m$ &
Comments \\
\hline

\multirow{3}{*}{$17^\circ$}
& \multirow{2}{*}{Periodic}
& Li \& Habbal
& $\approx 6.9$
& $\approx 6.15$
& $\approx 1.825$
& ---
& --- \\

&
& ALPS \& DIS-K
& $\approx 7.52$
& $\approx 6.71$
& $\approx 2.62$
& $\approx 43\%$ higher $\gamma_m$
& Extended unstable range \\

& Aperiodic
& Both
& Transition at $\approx 1.865$
& $\approx 1.65$
& $\approx 0.43$
& Very similar
& ---\\

\hline

\multirow{3}{*}{$20^\circ$}
& \multirow{2}{*}{Periodic}
& Li \& Habbal
& $\approx 7.21$
& $\approx 6.3$
& $\approx 1.95$
& ---
& --- \\

&
& ALPS \& DIS-K
& $\approx 8.3$
& $\approx 7.17$
& $\approx 3.48$
& $\approx 78\%$ higher $\gamma_m$
& Extended unstable range \\

& Aperiodic
& Both
& Transition at $\approx 4.04$
& $\approx 3.58$
& $\approx 1.49$
& Very similar
& --- \\

\hline

\multirow{4}{*}{$22^\circ$}
& \multirow{2}{*}{Periodic}
& Li \& Habbal
& $\approx 7.41$
& $\approx 6.35$
& $\approx 1.9$
& ---
& --- \\

&
& ALPS \& DIS-K
& $\approx 8.95$
& $\approx 7.54$
& $\approx 4.12$
& $\approx 117\%$ higher $\gamma_m$
& Periodic branch slightly dominant \\

& \multirow{2}{*}{Aperiodic}
& Li \& Habbal
& Transition at $\approx 5.80$
& $\approx 5.19$
& $\approx 2.85$
& ---
& Aperiodic branch dominant \\

&
& ALPS \& DIS-K
& Transition at $\approx 5.437$
& $\approx 4.83$
& $\approx 3.30$
& $\approx 16\%$ higher $\gamma_m$
& 
\\

\hline

\multirow{4}{*}{$24^\circ$}
& \multirow{2}{*}{Periodic}
& Li \& Habbal
& $\approx 7.61$
& Peak not fully developed
& $\approx 1.1$
& ---
& Weak periodic branch  \\

&
& ALPS \& DIS-K
& $\approx 9.74$
& $\approx 7.94$
& $\approx 4.77$
& $\approx 333\%$ higher $\gamma_m$
& Broad periodic peak \\

& \multirow{2}{*}{Aperiodic}
& Li \& Habbal
& Transition at $\approx 7.2$
& $\approx 6.25$
& $\approx 7.35$
& ---
& Strongly dominant \\

&
& ALPS \& DIS-K
& Transition at $\approx 6.517$
& $\approx 5.85$
& $\approx 6.09$
& $\approx 17\%$ lower $\gamma_m$
& Moderately dominant \\

\hline

\multirow{2}{*}{$70^\circ$}
& \multirow{2}{*}{Aperiodic}
& Li \& Habbal
& $\approx 34.4$
& $\approx 25.3$
& $\approx 183$
& ---
& --- \\

&
& ALPS \& DIS-K
& $\approx 36.3$
& $\approx 26.0$
& $\approx 192$
& $\approx 5\%$ higher $\gamma_m$
& Overall similar  \\

\hline

\multirow{2}{*}{$75^\circ$}
& \multirow{2}{*}{Aperiodic}
& Li \& Habbal
& $\approx 34.8$
& $\approx 26.9$
& $\approx 119$
& ---
& --- \\

&
& ALPS \& DIS-K
& $\approx 40.2$
& $\approx 29.7$
& $\approx 150$
& $\approx 26\%$ higher $\gamma_m$
& $\approx 15.5\%$ wider instability range \\

\hline

\multirow{2}{*}{$80^\circ$}
& \multirow{2}{*}{Aperiodic}
& Li \& Habbal
& $\approx 23.7$
& $\approx 19.5$
& $\approx 39$
& ---
& --- \\

&
& ALPS \& DIS-K
& $\approx 26.0$
& $\approx 21.0$
& $\approx 41$
& $\approx 5\%$ higher $\gamma_m$
& $\approx 10\%$ wider instability range \\

\hline
\hline
\end{tabular}
}
\end{table*}

\section{Validation. OPFHI with Maxwellian and SKD}\label{App_Validation_OPFHI}

\begin{figure}\centering
    \includegraphics[width=0.95\textwidth]{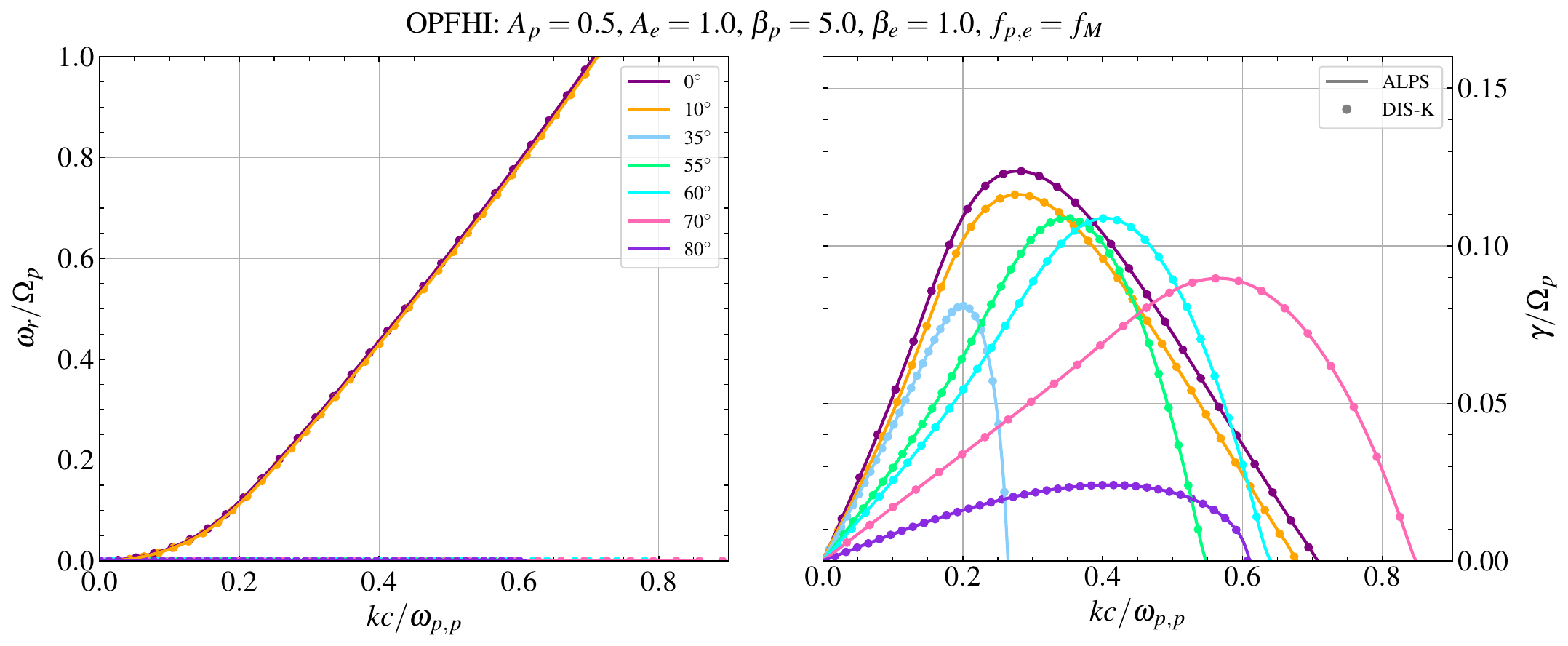}
    \vspace{2pt}
    
    \includegraphics[width=0.95\textwidth]{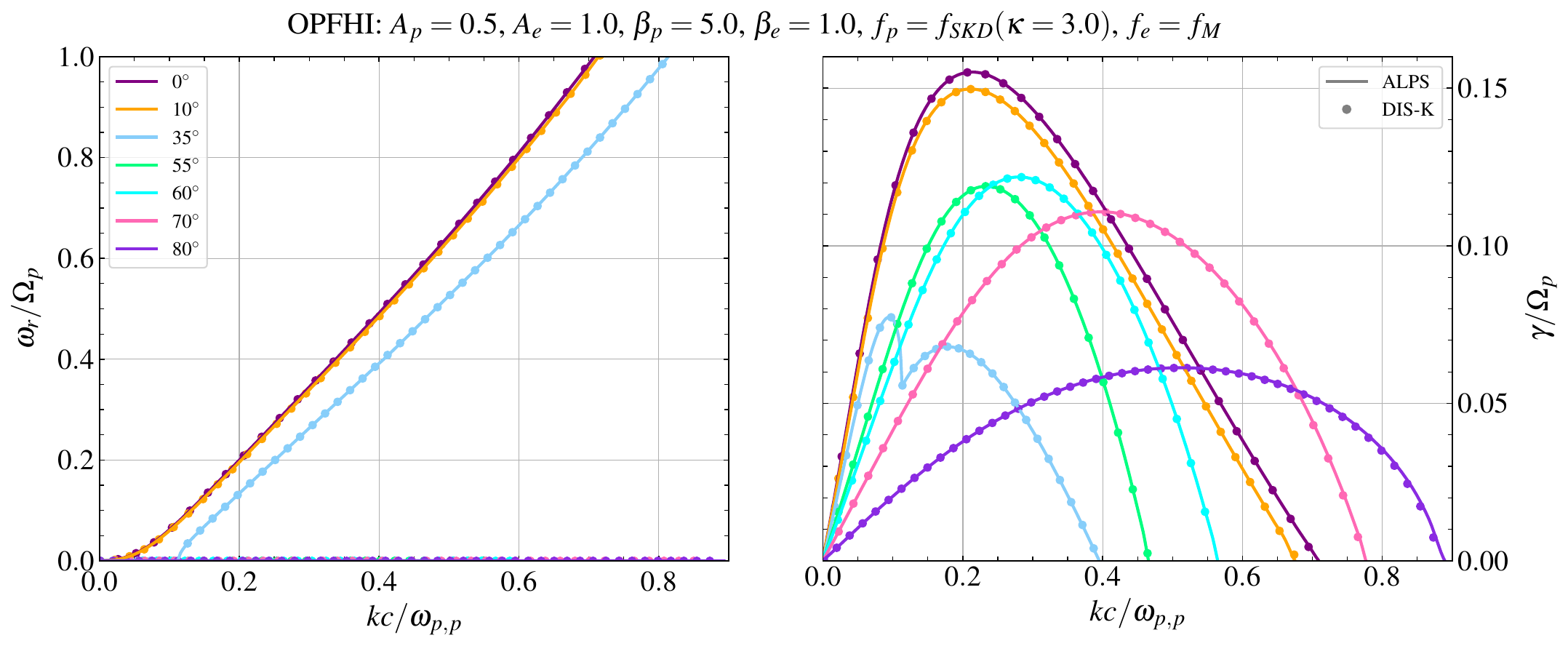}
    \caption{\textit{Normalized wave frequencies (left) and normalized growth rates (right) as functions of the normalized wave number for the OPFHI at different propagation angles. 
    Results obtained with ALPS (solid lines) and DIS-K (dots). Top: Maxwellian distribution. Bottom: SKD with $\kappa=3.0$.}}
    \label{Oblique_PFHI_validation}
\end{figure}

\begin{figure}
    \centering
    \includegraphics[width=0.95\textwidth]{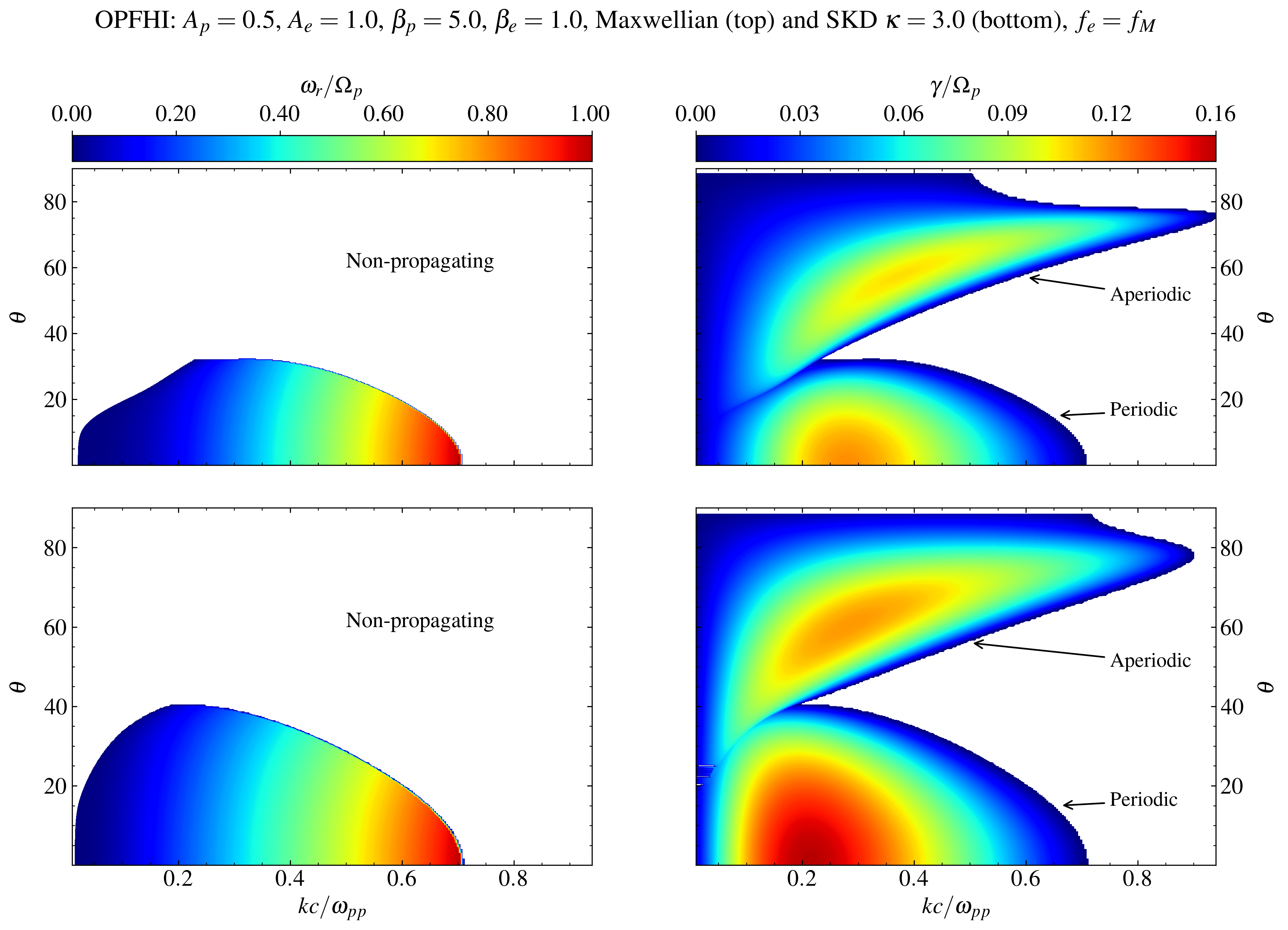}
    \caption{\textit{Contour plot of the OPFHI normalized frequency (left panel) and growth rate (right panel) in the $k$-$\theta$-plane for the Maxwellian case (top) and SKD case with $\kappa=3.0$ (bottom), obtained with DIS-K.}} \label{plot_Map_PFHI} 
\end{figure}

Inspired by \cite{HellingerMatsumoto_JGR_2000}, we use Maxwellian protons with $\beta_{p}=5$, and $T_{p,\perp}/T_{p,\|}=0.5$, while electrons are described by an isotropic Maxwellian with $\beta_{e}=1$. We extend this setup by using an SKD with $\kappa = 3.0$ for the electrons. The results for different propagation angles are presented in Figure \ref{Oblique_PFHI_validation}, with the normalized frequency on the left-hand side and the normalized growth rate on the right-hand side (top panel: Maxwellian, bottom panel: SKD $\kappa = 3.0$). The agreement between the two solvers is again very good (less than $0.1\%$ deviation for the Maxwellian cases and less than $0.4\%$ deviation for the SKD cases in terms of $\gamma_m$ and $k_m$) and the findings (for the Maxwellian case) are very similar to the ones reported by \cite{HellingerMatsumoto_JGR_2000}. Contour plots of the OPFHI normalized frequency (left panel) and growth rate (right panel) in the $k$-$\theta$-plane for the Maxwellian case (top) and for the SKD with $\kappa =3.0$ (bottom) are shown in Figure \ref{plot_Map_PFHI}, revealing that for both Maxwellian and SKD the maximum growth rate is reached for parallel propagation. For a detailed discussion about the OPFHI, we refer the reader to \cite{HellingerMatsumoto_JGR_2000} and \cite{Maneva_2016APJ}. 

\section{Maximum growth rates for OEFHI and OPFHI}
The maximum growth rates $\gamma_{m}/\Omega_p$  and the corresponding wave numbers $k_m c/\omega_{pp}$ for the different cases are shown in Table \ref{tab:max_gamma_oefhi} for the OEFHI and in Table \ref{tab:max_gamma_opfhi} for the OPFHI. 

\begin{table}
\centering
\caption{(O)EFHI: Maximum growth rate $\gamma_{m}$ and corresponding wavenumber $k_{m}$
for Maxwellian, SKD and different RKDs $(\kappa, \alpha)$ and propagation angles $\theta$. 
Modes are classified as periodic ($\omega_r \neq 0$) or aperiodic ($\omega_r = 0$). 
}
\label{tab:max_gamma_oefhi}
\setlength{\tabcolsep}{10pt} 
\renewcommand{\arraystretch}{0.85}
\begin{tabular}{c c @{\hspace{12pt}} c c @{\hspace{12pt}} c c}
\hline
\textbf{$\theta$} & \textbf{Distribution} 
& \multicolumn{2}{c}{\textbf{Periodic} ($\omega_r \neq 0$)} 
& \multicolumn{2}{c}{\textbf{Aperiodic} ($\omega_r = 0$)} \\
 &  & $k_{m}c/\omega_{pp}$ & $\gamma_{m}/\Omega_p$ 
    & $k_{m}c/\omega_{pp}$ & $\gamma_{m}/\Omega_p$ \\
\hline
\multirow{6}{*}{$0^\circ$}
& Maxwellian & 6.36 & 0.550 & -- & -- \\
& SKD (3.0) & 1.60 & 0.678 & -- & -- \\
& RKD (3.0, 0.15) & 2.20 & 0.734 & -- & -- \\
& RKD (2.0, 0.10) & 1.25 & 0.777 & -- & -- \\
& RKD (1.5, 0.10) & 0.92 & 0.818 & -- & -- \\
& RKD (1.0, 0.10) & 0.61 & 0.861 & -- & -- \\
\hline
\multirow{6}{*}{$10^\circ$}
& Maxwellian & 6.68 & 0.576 & 4.02 & 0.516 \\
& SKD (3.0) & 1.69 & 0.673 &  0.36  &  0.239\\
& RKD (3.0, 0.15) & 2.35 & 0.748 & 0.40 & 0.203 \\
& RKD (2.0, 0.10) & 1.30 & 0.781 & -- & -- \\
& RKD (1.5, 0.10) & 0.96 & 0.821 & -- & -- \\
& RKD (1.0, 0.10) & 0.63  &  0.862 & -- & -- \\
\hline
\multirow{6}{*}{$16^\circ$}
& Maxwellian &  7.13  & 0.617 & 6.12 & 2.971 \\
& SKD (3.0) & 1.84 & 0.669 & 0.40 & 0.251 \\
& RKD (3.0, 0.15) & 2.59 & 0.780 & 0.46 & 0.214 \\
& RKD (2.0, 0.10) & 1.41 &  0.791 & 0.30 & 0.300 \\
& RKD (1.5, 0.10) & 1.04 & 0.832 & 0.17 &  0.255 \\
& RKD (1.0, 0.10) & 0.67 & 0.866 & -- & -- \\
\hline
\multirow{6}{*}{$30^\circ$}
& Maxwellian & -- & -- & 9.59 &  21.15 \\
& SKD (3.0) & 2.70 & 0.717 & 1.98 & 1.167 \\
& RKD (3.0, 0.15) & -- & -- & 3.52 & 3.564 \\
& RKD (2.0, 0.10) & 1.95 & 0.893 & 0.65 &  0.463 \\
& RKD (1.5, 0.10) &  1.48  &  0.959 & 0.42 & 0.458 \\
& RKD (1.0, 0.10) & 1.04 &  0.996 & 0.25 & 0.430 \\
\hline
\multirow{6}{*}{$60^\circ$}
& Maxwellian & -- & -- & 20.8 & 113.6 \\
& SKD (3.0) & -- & -- & 16.3 & 126.5 \\
& RKD (3.0, 0.15) & -- & -- & 16.6 & 123.3 \\
& RKD (2.0, 0.10) & -- & -- & 15.3 & 128.4 \\
& RKD (1.5, 0.10) & -- & -- & 14.4 & 131.6 \\
& RKD (1.0, 0.10) & -- & -- & 13.2 & 138.7 \\
\hline
\multirow{6}{*}{$70^\circ$}
& Maxwellian & -- & -- & 27.7 & 104.6 \\
& SKD (3.0) & -- & -- & 21.7 & 189.8 \\
& RKD (3.0, 0.15) & -- & -- & 21.1 & 178.0 \\
& RKD (2.0, 0.10) & -- & -- & 20.6 & 211.3 \\
& RKD (1.5, 0.10) & -- & -- & 19.8 & 236.0 \\
& RKD (1.0, 0.10) & -- & -- & 18.7 & 281.1 \\
\hline
\multirow{6}{*}{$75^\circ$}
& Maxwellian & -- & -- & -- & -- \\
& SKD (3.0) & -- & -- & 24.4 & 191.6 \\
& RKD (3.0, 0.15) & -- & -- & 24.7 & 173.5 \\
& RKD (2.0, 0.10) & -- & -- & 23.4 & 228.2 \\
& RKD (1.5, 0.10) & -- & -- & 22.5 & 266.9 \\
& RKD (1.0, 0.10) & -- & -- & 21.2 & 335.4 \\
\hline
\multirow{6}{*}{$80^\circ$}
& Maxwellian & -- & -- & -- & -- \\
& SKD (3.0) & -- & -- & 25.4 & 134.2 \\
& RKD (3.0, 0.15) & -- & -- & 25.0 & 109.7 \\
& RKD (2.0, 0.10) & -- & -- & 25.4 & 188.7 \\
& RKD (1.5, 0.10) & -- & -- & 24.8 & 243.8 \\
& RKD (1.0, 0.10) & -- & -- & 23.6 & 339.2 \\
\end{tabular}
\end{table}

\begin{table}
\centering
\caption{(O)PFHI: Maximum growth rate $\gamma_{m}$ and corresponding wavenumber $k_{m}$
for Maxwellian, SKD and different RKDs $(\kappa, \alpha)$ and propagation angles $\theta$. 
Modes are classified as periodic ($\omega_r \neq 0$) or aperiodic ($\omega_r = 0$). 
}
\label{tab:max_gamma_opfhi}
\setlength{\tabcolsep}{10pt} 
\renewcommand{\arraystretch}{0.95}
\begin{tabular}{c c @{\hspace{12pt}} c c @{\hspace{12pt}} c c}
\hline
\textbf{$\theta$} & \textbf{Distribution} 
& \multicolumn{2}{c}{\textbf{Periodic} ($\omega_r \neq 0$)} 
& \multicolumn{2}{c}{\textbf{Aperiodic} ($\omega_r = 0$)} \\
 &  & $k_{m}c/\omega_{pp}$ & $\gamma_{m}/\Omega_p$ 
    & $k_{m}c/\omega_{pp}$ & $\gamma_{m}/\Omega_p$ \\
\hline
\multirow{6}{*}{$0^\circ$}
& Maxwellian & 0.279 & 0.123 & -- & -- \\
& SKD (3.0) & 0.213 & 0.155 & -- & -- \\
& RKD (3.0, 0.15) & 0.219 & 0.152 & -- & -- \\
& RKD (2.0, 0.1) & 0.191 & 0.160 & -- & -- \\
& RKD (1.5, 0.1) & 0.168 & 0.163 & -- & -- \\
& RKD (1.0, 0.1) & 0.131 & 0.169 & -- & -- \\
\hline
\multirow{6}{*}{$10^\circ$}
& Maxwellian & 0.278 & 0.116 & -- & -- \\
& SKD (3.0) & 0.210 & 0.149 & -- & -- \\
& RKD (3.0, 0.15) & 0.216 & 0.147 & -- & -- \\
& RKD (2.0, 0.1) & 0.188 & 0.155 & -- & -- \\
& RKD (1.5, 0.1) & 0.166 & 0.158 & -- & -- \\
& RKD (1.0, 0.1) & 0.128 & 0.164 & -- & -- \\
\hline
\multirow{6}{*}{$35^\circ$}
& Maxwellian & -- & -- & 0.200 & 0.0809 \\
& SKD (3.0) & 0.179 & 0.0679 & 0.0947 & 0.0774 \\
& RKD (3.0, 0.15) & 0.188  & 0.0614 & 0.108  & 0.0816 \\
& RKD (2.0, 0.1) & 0.150 & 0.0843 & 0.0673 & 0.0733 \\
& RKD (1.5, 0.1) & 0.128 & 0.0954 & 0.0483 & 0.0707 \\
& RKD (1.0, 0.1) & 0.097 & 0.109 & 0.0349 & 0.0743 \\
\hline
\multirow{6}{*}{$55^\circ$}
& Maxwellian & -- & -- & 0.347 & 0.109 \\
& SKD (3.0) & -- & -- & 0.233 & 0.119 \\
& RKD (3.0, 0.15) & -- & -- & 0.244 & 0.118 \\
& RKD (2.0, 0.1) & -- & -- & 0.200 & 0.120 \\
& RKD (1.5, 0.1) & -- & -- & 0.171 & 0.120 \\
& RKD (1.0, 0.1) & -- & -- & 0.126 & 0.122 \\
\hline
\multirow{6}{*}{$60^\circ$}
& Maxwellian & -- & -- & 0.403 & 0.108 \\
& SKD (3.0) & -- & -- & 0.280 & 0.121 \\
& RKD (3.0, 0.15) & -- & -- & 0.291  & 0.120 \\
& RKD (2.0, 0.1) & -- & -- & 0.245 & 0.124 \\
& RKD (1.5, 0.1) & -- & -- & 0.215 & 0.126 \\
& RKD (1.0, 0.1) & -- & -- & 0.169 & 0.129 \\
\hline
\multirow{6}{*}{$70^\circ$}
& Maxwellian & -- & -- & 0.565 & 0.0897 \\
& SKD (3.0) & -- & -- & 0.397 & 0.110 \\
& RKD (3.0, 0.15) & -- & -- & 0.410  & 0.108 \\
& RKD (2.0, 0.1) & -- & -- & 0.354 & 0.115 \\
& RKD (1.5, 0.1) & -- & -- & 0.318 & 0.119 \\
& RKD (1.0, 0.1) & -- & -- & 0.262 & 0.123 \\
\hline
\multirow{6}{*}{$80^\circ$}
& Maxwellian & -- & -- & 0.409 & 0.0241 \\
& SKD (3.0) & -- & -- & 0.512  & 0.0613 \\
& RKD (3.0, 0.15) & -- & -- & 0.519  & 0.0576 \\
& RKD (2.0, 0.1) & -- & -- & 0.483 & 0.0687 \\
& RKD (1.5, 0.1) & -- & -- & 0.450 & 0.0741 \\
& RKD (1.0, 0.1) & -- & -- & 0.397 & 0.0793 \\
\hline
\end{tabular}
\end{table}

\bibliographystyle{jpp}

\bibliography{jpp-instructions}

\end{document}